# ciftiTools: A package for reading, writing, visualizing, and manipulating CIFTI files in R

Damon Pham [1], John Muschelli [2], Amanda Mejia [1]

1. Department of Statistics, Indiana University, USA; 2. Department of Biostatistics, Johns Hopkins Bloomberg School of Public Health, USA

**Abstract**

There is significant interest in adopting surface- and grayordinate-based analysis of MR data for a number of reasons, including improved whole-cortex visualization, the ability to perform surface smoothing to avoid issues associated with volumetric smoothing, improved inter-subject alignment, and reduced dimensionality. The CIFTI grayordinate file format introduced by the Human Connectome Project further advances grayordinate-based analysis by combining gray matter data from the left and right cortical hemispheres with gray matter data from the subcortex and cerebellum into a single file. Analyses performed in grayordinate space are well-suited to leverage information shared across the brain and across subjects through both traditional analysis techniques and more advanced statistical methods, including Bayesian methods. The R statistical environment facilitates use of advanced statistical techniques, yet little support for grayordinates analysis has been previously available in R. Indeed, few comprehensive programmatic tools for working with CIFTI files have been available in any language. Here, we present the ciftiTools R package, which provides a unified environment for reading, writing, visualizing, and manipulating CIFTI files and related data formats. We illustrate ciftiTools' convenient and user-friendly suite of tools for working with grayordinates and surface geometry data in R, and we describe how ciftiTools is being utilized to advance the statistical analysis of grayordinate-based functional MRI data.

# 1 Introduction

Surface-based analysis of magnetic resonance (MR) data was pioneered by Freesurfer (Fischl 2012) and further popularized by the Human Connectome Project (HCP) (Van Essen et al. 2013). There is significant interest in adopting surface-based analysis for a number of reasons, including improved whole-cortex visualization (Fischl, Sereno, and Dale 1999; Van Essen 2012), mitigating problems associated with volumetric smoothing such as reduced sensitivity and specificity (Coalson, Van Essen, and Glasser 2018; Brodoehl et al. 2020), improved inter-subject alignment of cortical folding patterns and functional areas (Fischl et al. 1999; Anticevic et al. 2008; Glasser et al. 2016), and reduced dimensionality.

Until fairly recently, surface-based MR data were only stored in data formats that exclude relevant subcortical and cerebellar gray matter regions of the brain. For example, GIFTI files are a type of surface format that can contain either metric data (e.g. functional MRI [fMRI] timeseries), label data (e.g. a parcellation) or surface geometry data for a particular hemisphere. Incorporating subcortical and/or cerebellar gray matter in an analysis would require pairing the surface data with volumetric data stored in a traditional NIFTI file. To address this limitation and to advance and promote the adoption of "grayordinates" MR data—a term referring collectively to the cortical, subcortical, and cerebellar gray matter —the HCP introduced the CIFTI file format (Glasser et al. 2013). "CIFTI" is an acronym for the "Connectivity Informatics Technology Initiative." (The "N" in "NIFTI" stands for "Neuroscience" and the "G" in "GIFTI" stands for "Geometry.") CIFTI files are a type of NIFTI file with a numeric matrix for all the grayordinates data, paired with an XML header containing information on the size and type of the data, column names, and brain structure labels. CIFTI files are thus a single data representation for all grayordinates. This combined format provides a convenient and compact file structure for users to perform analysis across all gray matter areas of the brain.

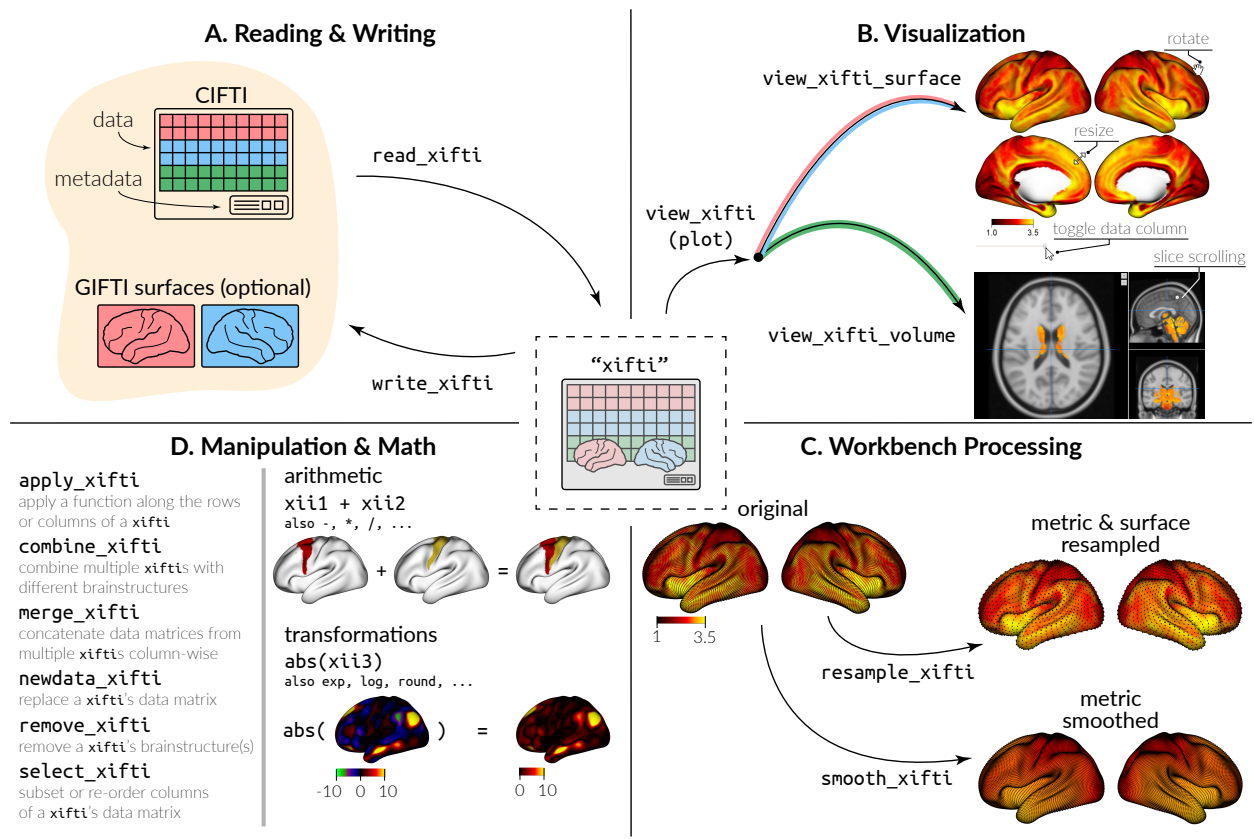

Figure 1: **Summary of `ciftiTools` functionality.** `ciftiTools` supports reading and writing (A), visualization (B), Workbench processing (C), and data manipulation and math (D) for CIFTI- and GIFTI-format data.

|  | Language | Read | | Write | | Plot | | Manipulate | |
|---|---|---|---|---|---|---|---|---|---|
|  |  | CIFTI | GIFTI | CIFTI | GIFTI | Cortex | Subcortex | CIFTI | GIFTI |
| ciftiTools | R* | 3 | ✓ | 3 | ✓ | ✓ | ✓ | 3 | ✓ |
| cifti | R | ✓ | | | | | | | |
| gifti | R | | ✓ | | ✓ | | | | |
| freesurfer-formats, fsbrain | R | ✓ | ✓ | | ✓ | ✓ | | | |
| ggseg, ggseg3d, ggsegExtra | R* | | | | | ✓ | ✓ | | |
| ciftify | Python* | | | ✓ | | ✓ | ✓ | | |
| NiBabel | Python | ✓ | ✓ | ✓ | ✓ | | | ✓ | |
| nilearn | Python | | ✓ | | | ✓ | ✓ | | |
| hcp-utils | Python | | | | | ✓ | ✓ | ✓ | |
| cifti-matlab | MATLAB | ✓ | | ✓ | | | | ✓ | |
| GIfTI | MATLAB | | ✓ | | ✓ | ✓ | | | |

**Table 1: Summary of comparison between programmatic tools for CIFTI and GIFTI files.** Checkmarks are assigned if the tool (rows) directly enables a user to complete the task (columns). A "3" for CIFTI functionalities in `ciftiTools` indicates that only three CIFTI intents are supported: "dtseries," "dscalar," and "dlabel." Moreover, tools receiving checkmarks in the same column may enable the task to different extents. For example, the manipulation functions in `hcp-utils` support normalizing data values and applying a parcellation, whereas those in `cifti-matlab` support data and metadata replacement. While `ciftiTools` is designed to support all listed tasks, other tools are not meant to support all these tasks on their own. For example, `hcp-utils` is designed to be used in conjunction with `NiBabel` and `nilearn`. Lastly, note that some tools have external dependencies: `ciftiTools` also requires the Connectome Workbench for most of its functionality, `ggsegExtra` requires FreeSurfer for reading in new parcellations, and `ciftify` uses FSL, FreeSurfer and the Connectome Workbench to convert and manipulate CIFTI files. An asterisk has been added to the "Language" column for these tools to indicate the need for additional software.

Grayordinate-based analysis of MR data also facilitates leveraging information shared across (1) individuals, since spatial alignment of functional areas across subjects is improved, and (2) the brain, since distances along the cortex and within specific gray matter structures tend to be inversely related to similarity of functional and structural features. Traditional MR analysis techniques benefit from these properties when adapted to grayordinate space, such as group averaging, geodesic surface smoothing and parcel-constrained volumetric smoothing. Accuracy and power can be further improved through Bayesian statistical techniques, many of which are implemented in the R statistical environment (R Core Team 2021). Examples include Markov chain Monte Carlo (Martin, Quinn, and Park 2011), an interface to the Stan probabilistic programming language (Stan Development Team 2021), Bayesian hierarchical models (Plummer 2003), integrated nested Laplace approximation (INLA) for fast and accurate posterior estimation (Rue, Martino, and Chopin 2009), spatial process priors suitable for surface-based modeling (Lindgren and Rue 2015), and excursion set approaches useful for performing joint inference across the cortex to leverage spatial information (Bolin and Lindgren 2016). However, the ability to use these advanced statistical methods is currently limited by a lack of tools available for working with CIFTI-format and other surface-based neuroimaging data in R.

Here, we present the `ciftiTools` R package, which provides a user-friendly, high-level interface for accessing and analyzing CIFTI and GIFTI data. Specifically, `ciftiTools` is a suite of tools for reading, writing, visualizing, processing, and manipulating CIFTI data and compatible surface geometries (**Figure 1**). It supports three CIFTI intents or file types: "dtseries" (for timeseries data), "dscalar" (for structural, continuously-valued data), and "dlabel" (for parcellations, labels, or categorical data). It also supports GIFTI metric data (analogous to "dtseries" or "dscalar") and label data (analogous to "dlabel") for wider applicability to surface-based analysis. We introduce the object class `"xifti"`, which refers to the combination of metric or label data (traditionally encoded in CIFTI files or GIFTI metric or label files) with surface geometries encoded in GIFTI surface geometry files. This facilitates the visualization, processing, and analysis of surface data, since the data and its geometric organization are contained within a single object. Below, we describe the structure and functionality of the `ciftiTools` package (Section 2) and demonstrate its use in an example seed correlation analysis (Section 3). We also explain its relationship with other packages and applications (Section 4, summarized in **Table 1**) to illustrate that `ciftiTools` uniquely provides unified support for reading, writing, plotting, and manipulating CIFTI-format data.

# 2 Overview of ciftiTools structure and functionality

**Figure 1** illustrates the general use case for `ciftiTools`. A `"xifti"` object can be constructed from either a CIFTI file or a combination of GIFTI metric or label files and a NIFTI file. In either case, two GIFTI surface geometry files may also be included (Panel A). The `"xifti"` can subsequently be visualized (Panel B), processed and manipulated (Panels C and D), and written back to a file(s) (Panel A). The integration of surface geometry is convenient because surfaces are required for visualizing and smoothing the cortical data, since no cortical spatial information is included in CIFTI files. Subcortical voxel locations and brain structure labels, in contrast, are contained in the CIFTI XML.

In this section, we first describe the structure of the `"xifti"` object class, and then describe the functionality illustrated in each panel of **Figure 1** in greater detail. For the sake of brevity we will not catalog every function belonging to each panel, nor will we cover the functions which do not fit into any of the four panels. However, an organized list of the most commonly used functions can be accessed in Appendix A and is reproduced in R with the command `help(ciftiTools)`. A complete, alphabetical list of all `ciftiTools` functions can be accessed with the command `help(package="ciftiTools")`. For each function, more information can be accessed at its help page using the command `help(function_name)`, e.g. `help(read_xifti)`.

## 2.1 The "xifti" object class

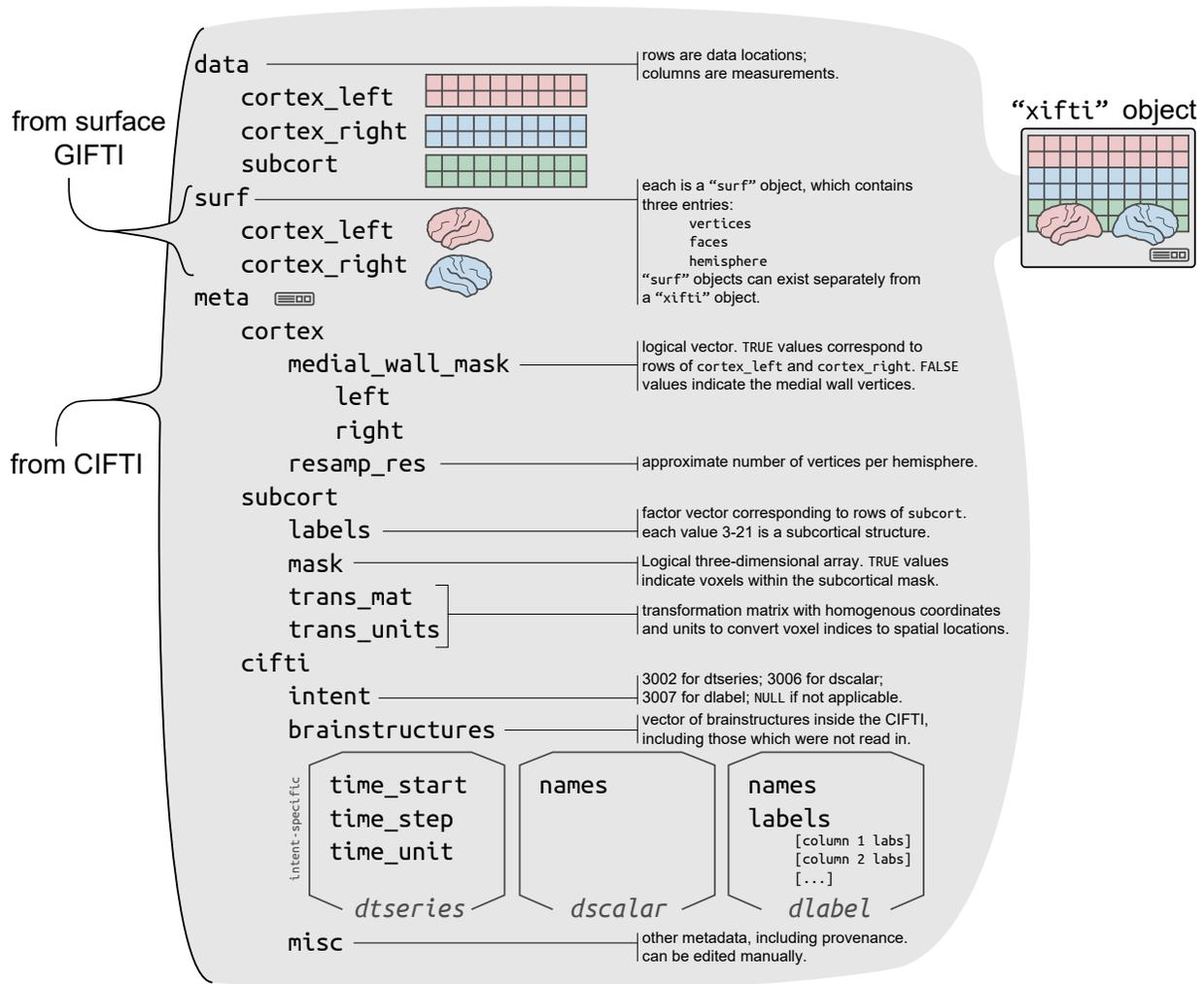

**Figure 2: The structure of a `"xifti"` object.** A `"xifti"` object can be constructed from CIFTI data and GIFTI surface geometry data using `read_xifti`. The object is a nested, ordered list, and any components that are absent will be `NULL`, for example if the CIFTI did not contain every brain structure or if no surfaces were provided. Using `read_xifti2` or `as.xifti`, it is also possible to construct a `"xifti"` object from GIFTI files which contain cortical data and/or a NIFTI file which contains subcortical data, rather than from a CIFTI file. In this case, the CIFTI metadata (intent, brain structures, intent-specific metadata, misc.) will be absent.

**Figure 2** illustrates the structure of the `"xifti"` object class. It is organized as a nested list with three entries at the first level: `data`, the data matrices for each brain structure; `surf`, the left and right cortical surface geometries; and `meta`, the various metadata including medial wall masks (see Appendix B for information about the medial wall), subcortical labels and mask, and CIFTI intent-specific information. If any entry is not present in the `"xifti"` it will have a value of `NULL`. Since a `"xifti"` is a list, its contents can be parsed with `$`. For example, `xii$meta$cifti$intent` returns the CIFTI intent of the `"xifti"` object `xii`.

The left and right cortical surface geometries are themselves objects of class `"surf"`. A `"surf"` object is a list with three entries: `vertices`, `faces`, and `hemisphere`. A `"surf"` object can also exist separately from any `"xifti"` object, allowing for direct visualization, resampling, and manipulation of surface geometry data in the absence of metric or label data.

The function `is.xifti` can be used to verify that an R object is a correctly-formatted `"xifti"` object. This validation is especially useful if the user has manually manipulated the data or metadata in a `"xifti"` and requires it to be correctly formatted, for example if they plan to write it to a file, visualize it, or further process it with `ciftiTools`.

Several S3 methods are implemented in `ciftiTools`. S3 methods control the behaviors of certain base R functions when applied to objects of a specific class. Currently, `"xifti"` objects have S3 methods for `summary`, `print`, `plot`, and several mathematical operations. For example, `plot(xii)` will invoke the functions `view_xifti_surface` and/or `view_xifti_volume`, as described in Section 2.3. `"surf"` objects have S3 methods for `summary`, `print`, and `plot`. For both `"xifti"` and `"surf"` objects, `summary` and `print` will display an overview of the object's contents. Other S3 methods for `"xifti"` and `"surf"` objects are described in the corresponding subsections below.

## 2.2 Reading and writing

**Figure 1A** shows how the `ciftiTools` function `read_xifti` imports CIFTI data into R as a `"xifti"` object. `read_xifti` may also import GIFTI surface geometry data alongside the CIFTI data. Any surfaces will be resampled to match the cortical data resolution in the CIFTI, if there is a difference in resolution. The flow of data is reversed when a `"xifti"` is written out with `write_xifti`: surfaces, if any, are written to GIFTI files, and everything else is written to a CIFTI file.

Two functions, `as.xifti` and `read_xifti2`, provide alternative methods for constructing a `"xifti"` object. `as.xifti` creates a `"xifti"` from R data matrices instead of a CIFTI file, allowing the user to store grayordinate data which may not originate from real neuroimaging, e.g. simulated data. For example, with `as.xifti` the user can simulate data using a number of powerful simulation tools in R, and then format the data as `"xifti"` grayordinate data. This `"xifti"` can then be analyzed with `ciftiTools`, or it can be written to a CIFTI file for analysis or visualization with methods and software outside of R. The other alternative function, `read_xifti2`, creates a `"xifti"` from GIFTI metric or label files instead of a CIFTI file. It is provided to allow for greater flexibility in analyses that are focused on the cortical surface specifically, which would not require the CIFTI format.

Additional functions related to reading and writing include `separate_xifti`, which splits a CIFTI file into GIFTI metric or label files for the cortical data and a NIFTI file for the subcortical data, and `info_xifti`, which reads in only the metadata of a CIFTI file. `write_xifti2` does the opposite of `read_xifti2`: it writes a `"xifti"` with surface data to GIFTI metric or label files. And for surfaces, `read_surf` and `write_surf` translate between GIFTI surface geometry files and `"surf"` objects.

A `"xifti"` object can also be written to an RDS data file using `saveRDS`, and read back in with `readRDS`, both base R functions. This allows R users to store CIFTI data and GIFTI surfaces in a single file. However, RDS files are generally not compatible with other programming platforms such as MATLAB or Python. Users interested in cross-platform compatibility should instead use `write_xifti` to generate CIFTI and GIFTI surface geometry files for subsequent use with MATLAB, Python, or other tools (see Section 4.4).

## 2.3 Visualization

**Figure 1B** illustrates how the grayordinates data in a `"xifti"` object can be visualized with `view_xifti`. Equivalently, the generic R function `plot` can be applied to a `"xifti"` object, since `view_xifti` is the `plot` S3 method for `"xifti"` objects. `view_xifti` will plot the cortical data with `view_xifti_surface` and the subcortical data with `view_xifti_volume`. If the `"xifti"` object contains both cortical and subcortical data, the two plots will share the same color scale, palette, and color scale limits. `view_xifti_surface` and `view_xifti_volume` can also both be called directly. The functions have similar interfaces: for example, they both can produce either interactive widgets or static images exportable to PNG files. However, they differ in a few notable ways, as described below.

The `view_xifti_surface` function produces a 3D rendering of the cortical surface data using the `rgl` package (Murdoch and Adler 2021). By default, the data are plotted on averaged "inflated" surfaces from the HCP included with `ciftiTools` (Glasser et al. 2013) unless other surfaces are contained in the `"xifti"` or provided in the optional function arguments. (The default "inflated" surfaces are shown in **Figure 1**; see Appendix D for a list of other surfaces included in `ciftiTools`). Additional arguments control the color scale type (sequential, diverging, or qualitative); palette and color scale limits; the hemisphere(s) and view(s) (medial and/or lateral) to show; the data column(s) to display; and whether the plot should be interactive or written to a PNG file(s).

If an interactive plot is created with `view_xifti_surface`, the user may click and drag to rotate the surfaces and scroll to resize them. The interactive plot will usually open in an OpenGL window (Shreiner 2009) but in specific cases will open in an HTML widget (Vaidyanathan et al. 2020). The help page `help(view_xifti_surface)` explains in depth when each is used, but in general, the HTML widget is used only if: multiple measurements are requested (since it can include a slider bar to control the measurement being displayed); the argument `widget` is set to `TRUE`; or if the computing platform is not compatible with OpenGL. All three output types (PNG files, OpenGL windows, and HTML widgets) can be embedded in HTML R Markdown documents, and the PNG files can be embedded in PDF R Markdown documents (see Appendix C). Note that for PNG files and HTML widgets, each plot first needs to be rendered in a temporary OpenGL window which will close automatically.

The `view_xifti_volume` function produces a 2D slicewise plot of the subcortical and cerebellar data overlaid on a 2 mm MNI structural template by default (Evans et al. 2012). It has the same arguments as `view_xifti_surface` for controlling the measurement(s) being plotted and the interactivity or file saving. The non-interactive plot is based on the `overlay` function from the `oro.nifti` package (Whitcher, Schmid, and Thornton 2011) and additionally has the same arguments for color scale type, palette, and color scale limits as `view_xifti_surface`. It also includes arguments to control the anatomical plane and slice indices being shown. The interactive plot is a wrapper to the namesake function of the `papayar` package (Muschelli 2016) and allows the user to scroll through slices along each anatomical axis and read the value at each voxel. The color scale, palette, and color scale limits can be adjusted within the GUI. The non-interactive plot can be exported to a PNG file and then embedded in R Markdown documents (see Appendix C).

`"surf"` objects can also be visualized independently of any `"xifti"` data using the function

`view_surf`. The `plot` S3 method for `"surf"` objects serves as an alias for `view_surf`. Like `view_xifti_surface`, `view_surf` is based on the `rgl` package (Murdoch and Adler 2021) and includes arguments to control the view(s) to show and whether the plot should be interactive or written to a PNG file.

`ciftiTools` also includes the function `view_comp` to assist users with displaying multiple PNG files at a time. `view_comp` could be used to display the data for multiple columns or `"xifti"` objects all in a grid, or to display corresponding cortical and subcortical plots side-by-side.

## 2.4 Processing

The Connectome Workbench is a suite of open-source software tools for neuroimaging data designed especially for HCP data (Marcus et al. 2011). It includes numerous command-line tools that can execute operations on CIFTI files and other data formats. The `ciftiTools` package provides a convenient interface for a few Workbench commands including two common processing operations, namely resampling and smoothing, as illustrated in **Figure 1C**. Both `"xifti"` objects and CIFTI files can be directly resampled and smoothed using `ciftiTools`, as described below. A full list of Workbench commands used by `ciftiTools` is in Section 4.1.1, and more interfaces to Workbench commands may be implemented in the future based on user needs.

To resample cortical data with `resample_xifti`, the user simply provides the `"xifti"` object or path to a CIFTI file, along with the target resolution specified in vertices per hemisphere. Other necessary files, such as the spherical geometries in-register with the target and original resolutions, will be generated automatically in a temporary directory. Note that the exact number of vertices after resampling may differ slightly from the target resolution due to the spherical mapping process by which resampling is performed. Also, note that the subcortex and cerebellum are never resampled. If surfaces are included in the `"xifti"` object or are specified as GIFTI files by the user, they will be resampled along with the CIFTI data matrix to the same target resolution. It is also possible to resample metric, label, and surface geometry GIFTI files with the function `resample_gifti` and `"surf"` objects with the function `resample_surf`.

Smoothing can be performed geodesically along the cortical surface and volumetrically for subcortical and cerebellar data using the function `smooth_cifti`. The user simply provides a `"xifti"` object or path to a CIFTI file and the full width at half maximum (FWHM) of the smoothing kernel. Cortical geodesic smoothing will be performed based on the default "inflated" surfaces unless other surfaces are included in the `"xifti"` or provided in the optional function arguments. By default, volumetric smoothing is constrained within each subcortical structure to avoid blurring across regional boundaries. It is also possible to smooth metric GIFTI files with the function `smooth_gifti`.

One consideration for both `resample_xifti` and `smooth_xifti` is that they will process a CIFTI file faster than its equivalent `"xifti"` object. This is because the Connectome Workbench commands operate on CIFTI files, so "under the hood" of both functions operating on a xifti object, that object must be written to a CIFTI file, resampled or smoothed, and then read back in. Writing a `"xifti"` object is trivially fast for CIFTI data with one or few measurements, but can be slow for a high-resolution CIFTI with many measurements, such as fMRI timeseries data. In the latter case, it is therefore more efficient to directly resample or smooth the CIFTI file before reading it in, if possible.

## 2.5 Math and manipulation

As illustrated in **Figure 1D**, `ciftiTools` provides several convenient functions for altering `"xifti"` objects while preserving a valid `"xifti"` object structure. These fall into two categories: manipulation and math. *Manipulation* functions explicitly alter one or multiple `"xifti"` objects. Several of these are listed on the left of panel D (see Appendix A for a full list with descriptions). They include functions to add or remove brain structures, subset the data matrix or combine it with others, etc. Manipulation functions reduce the need for users to manually edit individual brain structure entries within a `"xifti"` object while ensuring that the metadata is updated appropriately to preserve a valid `"xifti"` format. *Math* functions perform mathematical operations, such as arithmetic and univariate transformations, on just the data matrix of one or more `"xifti"` objects. These are base R functions that can be applied directly to `"xifti"` objects. For example, the data matrices of two `"xifti"` objects can be added together with the command `xii1 + xii2`; the data matrix of a single `"xifti"` object can be log-transformed using the command `log(xii1)`. `xii1 - xii1 == 0` returns a `"xifti"` with all values equal to 1, representing `TRUE`. All Math functions are implemented as S3 methods for `"xifti"` objects; as such, they are limited to fundamental arithmetic and univariate operations (see Appendix A for a full list with descriptions). To perform more advanced mathematical manipulation or analysis of `"xifti"` objects, we recommend that users extract the data matrix with `as.matrix`, make the desired changes, and then update the data matrix of the `"xifti"` object using the function `newdata_xifti`. For simple operations, though, the manipulation and math functions included in `ciftiTools` allow users to work with CIFTI data conveniently and with concise code, while providing the building blocks for users to create more involved processing routines specific to their needs.

## 3 Example

In this section, we will illustrate some of the `ciftiTools` functionality through a simple example analysis of computing and visualizing seed correlations for a resting-state fMRI (rfMRI) scan from the HCP (Van Essen et al. 2013) which has been pre-processed with ICA-FIX (Salimi-Khorshidi et al. 2014). Our seed will be the posterior cingulate cortex (PCC), a highly-connected node of the default mode network (DMN). We will use functional parcellations created by Schaefer et al. (2018) to define the PCC and other brain regions. The analysis is divided into four steps: (1) reading in and smoothing the fMRI data; (2) reading in the parcellation; (3) calculating seed correlations; and (4) visualizing and saving the results. An expanded demonstration which includes data cleaning, the subcortical data, and more functions from `ciftiTools` is available in the Supplementary Materials. Note that we will read in the rfMRI data from a subfolder `Data`, and we will write out plot image files to a subfolder `Plots`, so to run this code again these two directories would need to be created, and a rfMRI CIFTI file would need to be downloaded from the HCP and placed in `Data`.

### 3.1 Reading in and smoothing the fMRI data

We begin by loading `ciftiTools` and pointing to the Connectome Workbench folder.

```
library(ciftiTools)
ciftiTools.setOption("wb_path", "workbench")
```

```
## Using this Workbench path: 'workbench/bin_windows64/wb_command.exe'.
```

Now we can read in the CIFTI file containing the resting-state fMRI (rfMRI) data with `read_xifti`. By default, only the left and right cortex are read in: set `brainstructures = "all"` to also read in the subcortex. For simplicity, we will not include the subcortex in this analysis (see the expanded demonstration in the Supplementary Materials).

```
xii <- read_xifti("Data/rfMRI_FIX.dtseries.nii")
```

`xii` stores the CIFTI data as a `"xifti"` object. `summary`, or equivalently invoking the implicit `print` S3 method, will produce an overview of its contents:

```
xii
```

```
## ====CIFTI METADATA==================
## Intent:          3002 (dtseries)
## - time step      0.72 (seconds)
## - time start     0
## Measurements:    1200 columns
##
## ====BRAIN STRUCTURES================
## - left cortex    29696 data vertices
##                  2796 medial wall vertices (32492 total)
##
## - right cortex   29716 data vertices
##                  2776 medial wall vertices (32492 total)
```

The S3 method `as.matrix` will return the timeseries data in matrix form, with locations (vertices and/or voxels) along the rows and measurements (timepoints) along the columns. Functions like `dim` which coerce their inputs to data matrices will behave accordingly:

```
dim(xii)
```

```
## [1] 59412   1200
```

Now we will spatially smooth the data with `smooth_xifti` using a geodesic Gaussian smoother. We will use the default 5 mm FWHM width, but other levels of smoothing are possible using the `surf_FWHM` argument to `smooth_xifti`. As suggested at the end of Section 2.4, we smooth the original CIFTI file and then read it in, rather than smoothing `xii` which would be slower.

```
smooth_xifti("Data/rfMRI_FIX.dtseries.nii", "Data/rfMRI_FIX_sm.dtseries.nii")
xii_sm <- read_xifti("Data/rfMRI_FIX_sm.dtseries.nii")
```

We can plot the first columns of `xii` and `xii_sm` to get an idea of the data scale, range, and effect of smoothing. `plot` will create a 3D visualization of the cortical data by invoking

`view_xifti_surface`, as described above in Section 2.3. There are several options for displaying the plots (see Appendix C). Here we will plot each `"xifti"` separately, save the two plots to PNG files, and then composite the PNG files together with `view_comp`.

```
plot(xii, title="Original", fname="Plots/rfMRI.png")
```

```
## `zlim` not provided: using color range 0 - 13500 (data limits: 1160 - 17000).
```

```
plot(xii_sm, title="Smoothed", fname="Plots/rfMRI_sm.png")
```

```
## `zlim` not provided: using color range 0 - 13300 (data limits: 2600 - 15100).
```

```
view_comp(c("Plots/rfMRI.png", "Plots/rfMRI_sm.png"), ncol=2)
```

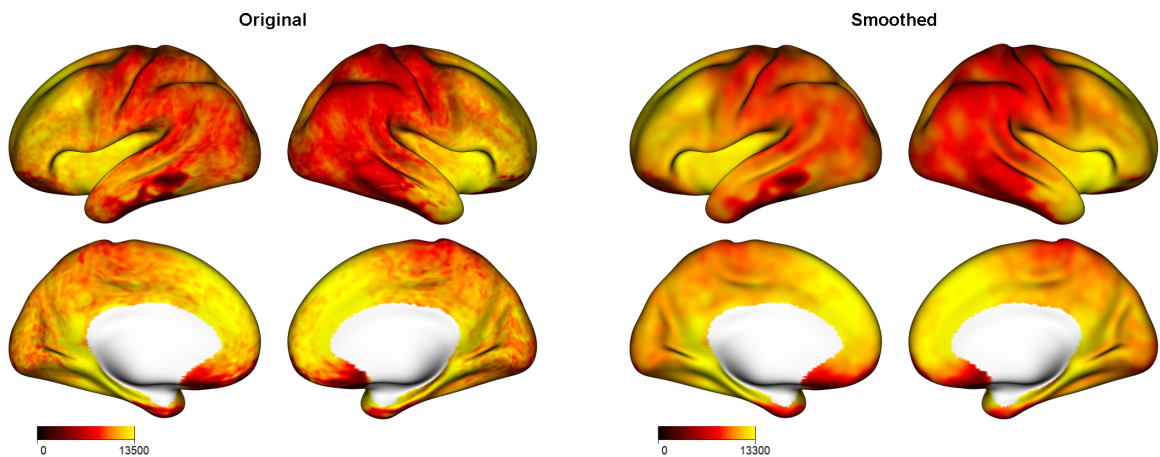

**Figure 3:** First fMRI volume before and after smoothing with smooth_xifti.

The color scale limits have been automatically determined based on the data, so they are close but slightly different. Nonetheless, the data on the right are clearly similar to the original except smoother across space.

## 3.2 Reading in the parcellation

Now we will load the parcellation. For convenience, ciftiTools includes a few commonly used parcellations, which can be read in using `load_parc` (refer to Appendix D) (Schaefer et al. 2018; Thomas Yeo et al. 2011). Here we will use the Schaefer parcellation with 400 parcels. Other parcellations in CIFTI or GIFTI format can be read in using `read_xifti` or `read_xifti2`, respectively.

```
parc <- load_parc("Schaefer_400")
```

Again, we can get an overview of the `"xifti"` object `parc` by using `summary`.

```
parc
```

```
## ====CIFTI METADATA==================
## Intent:          3007 (dlabel)
## - names          "parcels"
## Measurements:    1 column
##
## ====BRAIN STRUCTURES================
## - left cortex    32492 data vertices
##
## - right cortex   32492 data vertices
```

The parcellation includes data for the left and right cortical hemispheres at the same resolution of the rfMRI data (approx. 32,000 vertices per hemisphere). The parcellation has the "dlabel" intent which indicates that it contains label i.e. categorical data. Specifically, for each measurement (column) every data location is assigned an integer "key" which corresponds to a certain label and color. A table in the metadata gives the label and color of each key, for each measurement.

This file has only one measurement, `"parcels"`. Its keys range from 0 to 400, with 0 indicating the medial wall and 1-400 indicating a specific parcel. When visualizing the parcellation with `plot`, each parcel will be colored according to the label table colors by default. In this parcellation, parcels are grouped by network, and those in the same network are colored identically. Therefore, adjacent parcels in the same network are impossible to delineate by color. We can use the `border` argument of `view_xifti_surface` to draw outlines between vertices of different keys, allowing us to visualize both the networks, according to color, and the individual parcels, according to outlines.

We will also separately visualize parcel 14, which corresponds closely to the PCC. We use `transform_xifti` to create a mask of only that parcel. The full parcellation and the masked PCC parcel are shown side-by-side in **Figure 4**.

```
plot(parc, borders="black", title="Schaefer 400", fname="Plots/parc.png")

parc_PCC <- transform_xifti(parc, function(x){ ifelse(x == 14, 14, 0) })
plot(parc_PCC, borders="black", title="Parcel 14 (PCC)", fname="Plots/pPCC.png")

view_comp(c("Plots/parc.png", "Plots/pPCC.png"), ncol=2)
```

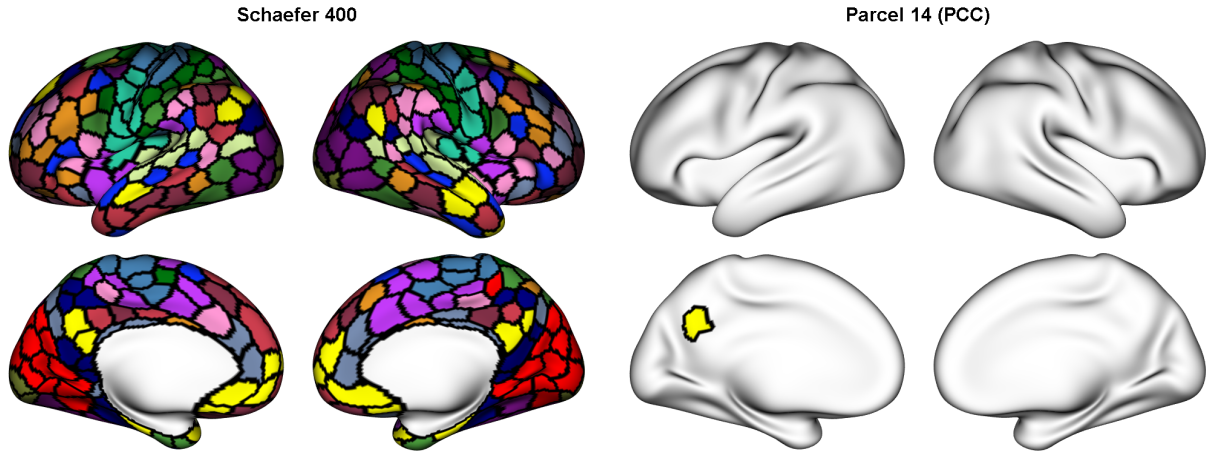

**Figure 4:** Schaefer parcellation and PCC parcel.

## 3.3 Computing seed correlations

Now that we have the smoothed rfMRI data and parcellation, we can compute the seed correlations. We begin by obtaining the mean timeseries of each parcel. This is easier done with the parcel data directly, rather than with the `parc "xifti"` object, so we will start by converting the parcellation to a vector. Then, we'll replace the medial wall vertices in `xii_sm` with `NA` data values so that the length of the parcellation vector, `parc_vec`, will match the number of cortical data vertices in `xii_sm`. (See Appendix B for information about the medial wall.)

```
parc_vec <- c(as.matrix(parc))
length(parc_vec)
```

```
## [1] 64984
```

```
xii_sm <- move_from_mwall(xii_sm, NA)
dim(xii_sm)
```

```
## [1] 64984   1200
```

Now for each parcel we will obtain the mean time series across its constituent vertices. Specifically, we will pre-allocate a matrix `xii_pmean`, and then for each parcel we will obtain the fMRI timeseries for all its vertices, take the average across space while ignoring medial wall values, and then save the resulting timeseries to a row of `xii_pmean`.

```
xii_mat <- as.matrix(xii_sm)
xii_pmean <- matrix(nrow=400, ncol=ncol(xii_mat))
for (p in 1:400) {
  data_p <- xii_mat[parc_vec==p,]
  xii_pmean[p, ] <- colMeans(data_p, na.rm=TRUE)
}
```

We then calculate the seed correlations between the PCC parcel and every other parcel.

```
seed_cor <- cor(t(xii_pmean))[,14]
```

## 3.4 Visualizing and saving the results

We can visualize the results by creating a new `"xifti"` with `newdata_xifti`. It will contain one measurement (column) in which the value of each vertex will be the seed correlation for the parcel it belongs to. Notice how the parcellation was converted from a `"xifti"` to a numeric matrix in the previous section with `as.matrix`, and after several mathematical operations, it's now being returned to a `"xifti"` with `newdata_xifti`. As suggested in Section 2.5, this is a workflow to use when it is not possible to obtain the result with only the operations built in to `ciftiTools`.

First we'll convert `parc_vec`, which gives the parcel key of each vertex, to `xii_seed`, which gives the seed correlation of the parcel each vertex belongs to. Then we will initialize a one-column `"xifti"` using `select_xifti` and replace its data with `xii_seed` using `newdata_xifti`.

```
xii_seed <- c(NA, seed_cor)[parc_vec + 1]
xii1 <- select_xifti(xii_sm, 1)
xii_seed <- newdata_xifti(xii1, xii_seed)
```

Now we can use the `"xifti"` S3 method `plot` again. We'll use it twice to demonstrate the ability of `view_xifti` to automatically choose an appropriate `zlim` (**Figure 5**).

```
plot(xii_seed, zlim=c(-1, 1), title="Set zlim", fname="Plots/seed.png")
plot(xii_seed, title="Auto zlim", fname="Plots/seed2.png")
```

```
## `zlim` not provided: using color range -0.345 - 0.345 (data limits: -0.374 - 1).
```

```
view_comp(c("Plots/seed.png", "Plots/seed2.png"), ncol=2)
```

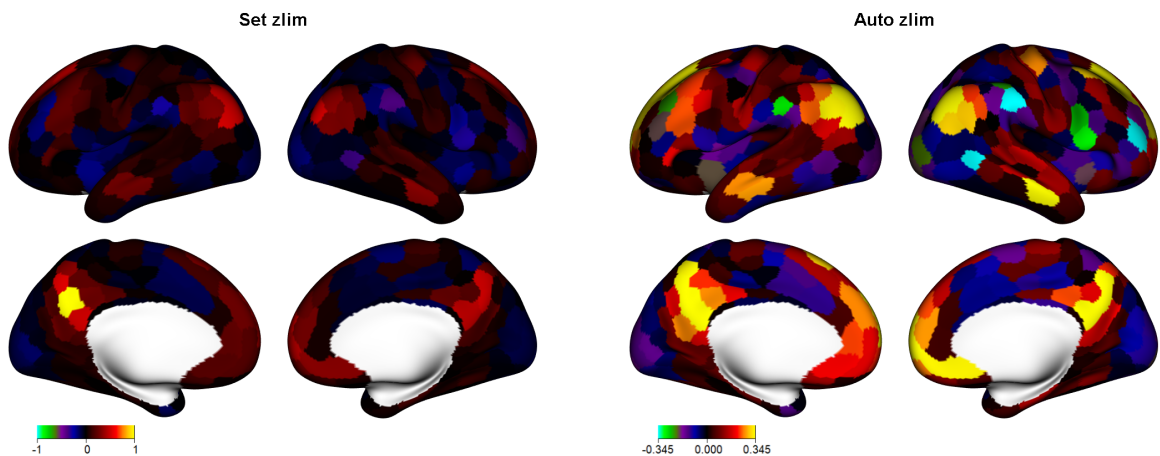

**Figure 5:** PCC seed correlation.

As expected, we find that most seed correlation values are positive or near zero, and the brain regions with the highest seed correlation are components of the DMN.

Lastly, we can save the result of our analysis with `write_xifti`. We will write the data to a "dscalar" CIFTI because it does not represent a timeseries ("dtseries") nor categorical/label data ("dlabel").

```
write_xifti(xii_seed, "Data/PCC_seedCor.dscalar.nii")
```

```
## Writing left cortex.
## Writing right cortex.
## Creating CIFTI file from separated components.
```

```
file.exists("Data/PCC_seedCor.dscalar.nii")
```

```
## [1] TRUE
```

# 4 Relationships with other tools and packages

In this section we describe a few key dependencies and reverse dependencies of `ciftiTools`, and we compare `ciftiTools` with similar tools in R as well as in other programming environments.

## 4.1 Dependencies of ciftiTools

`ciftiTools` has one external dependency, the Connectome Workbench, and several dependencies on other R packages.

### 4.1.1 Connectome Workbench

Several functions in `ciftiTools` depend on Connectome Workbench commands (Marcus et al. 2011), and are listed in **Table 2**. To access the full functionality of `ciftiTools`, including reading and writing files, the user must therefore have the Workbench installed in their computing environment, and the path to the Connectome Workbench must be indicated at the beginning of each session after loading the package as previously demonstrated in Section 3:

```
library(ciftiTools)
```

```
ciftiTools.setOption("wb_path", "/path/to/workbench"),
```

where `/path/to/workbench` is the location of either the Connectome Workbench folder or the executable file inside of it (`wb_command` on Mac and Linux or `wb_command.exe` on Windows). The path to the executable will be printed to indicate that it was set correctly; otherwise, a warning message will be displayed. If Workbench-dependent tasks are attempted prior to setting the Workbench path, an error message will be displayed.

| ciftiTools function | Workbench command(s) it interfaces with |
|---|---|
| info_cifti | -nifti-information |
| read_xifti | -cifti-convert, -metric-merge |
| write_xifti | -cifti-create-dense-timeseries, -cifti-create-dense-scalar, -cifti-create-label |
| separate_cifti | -cifti-separate |
| write_nifti | -volume-label-import |
| resample_xifti and resample_gifti | -cifti-separate, -metric-resample, -label-resample, -surface-resample, -surface-create-sphere, -surface-flip-lr, -set-structure, -cifti-create-dense-timeseries, -cifti-create-dense-scalar, cifti-create-label |
| resample_cifti_from_template | -cifti-resample |
| smooth_xifti | -cifti-smoothing, -cifti-change-mapping |
| smooth_gifti | -metric-smoothing |
| convert_xifti | -cifti-change-mapping |

Table 2: **Connectome Workbench commands used by `ciftiTools`.** Note that any Workbench command, including those not currently integrated in the package, can be executed from within R using the `ciftiTools` function `run_wb_cmd`, which conveniently appends the Workbench path to the beginning of the command. Workbench commands can also be executed within the R environment using the base R function `system`.

The `view_xifti_surface` function does not rely on Workbench commands, but was designed to emulate several aspects of visualization using the Workbench, including the layout and default color scales. Appendix E shows a comparison between a plot produced with `ciftiTools` and one produced with the Connectome Workbench for the same data. Similarly, some manipulation functions resemble Workbench commands but do not depend on them: for example, `merge_xifti` emulates `-cifti-merge`, `apply_xifti` is similar to `-cifti-reduce`, and `transform_xifti` is analogous to `-cifti-math`. These manipulation functions operate completely inside R, saving time by avoiding the need to write the `"xifti"` to a CIFTI file, process it with Workbench, and then read it back in, as is required for `resample_xifti` and `smooth_xifti`. `apply_xifti` and `transform_xifti` also benefit from being able to use any R function, including one created by the user, compared to the Workbench commands which are limited to predefined arguments.

#### 4.1.2 Other R packages

`ciftiTools` depends on several other R packages for working with GIFTI and NIFTI-format data. Specifically, the `gifti` package (Muschelli 2020) supports reading, writing, and parsing GIFTI files. The `RNifti` package (Clayden, Cox, and Jenkinson 2021) supports reading and writing NIFTI files, in particular subcortical data which has been separated from a CIFTI file. Cortical visualizations are based on the `rgl` package (Murdoch and Adler 2021). Subcortical visualizations are based on the `oro.nifti` package (Whitcher, Schmid, and Thornton 2011) for static plots and the `papayar` package (Muschelli 2016) for interactive plots. A full list of dependencies can be found in the package DESCRIPTION file.

## 4.2 R packages building on ciftiTools for grayordinates data analysis

Several R packages depend on `ciftiTools` to apply both traditional and more advanced statistical analysis techniques to fMRI in grayordinates format. The `BayesfMRI` package (Spencer, Pham, and Mejia 2021) can be used to perform general linear model (GLM) analysis on task fMRI data for identifying activations. It includes a spatial Bayesian approach designed for cortical surface analysis (Mejia, Yue, et al. 2020) in addition to the classical massive univariate approach (Friston et al. 1994). The package `templateICAr` (Mejia 2021) includes routines for group independent component analysis (ICA) and estimation of corresponding subject-level independent components (ICs) via dual regression (Beckmann et al. 2009) or template ICA (Mejia, Nebel, et al. 2020). Lastly, `fMRIscrub` (Pham, Muschelli, et al. 2021) includes routines for data-driven scrubbing (Smyser, Snyder, and Neil 2011; Mejia et al. 2017; Afyouni and Nichols 2018; Pham, McDonald, et al. 2021), motion scrubbing (Power et al. 2014), and denosing strategies such as anatomical CompCor (Behzadi et al. 2007).

## 4.3 Comparison with other R packages for working with CIFTI files

As listed in **Table 1**, a few R packages also support CIFTI files, to varying extents. Below we describe each package in more detail and compare its functionality with that of `ciftiTools`.

### 4.3.1 The cifti R package

The `cifti` R package reads in CIFTI files with any intent (Muschelli 2018). It does not currently support writing or visualizing CIFTI files. In contrast, `ciftiTools` was designed to work specifically with the "dtseries," "dscalar," and "dlabel" intents. Given this more limited scope, `ciftiTools` aims to provide easy access and analysis of CIFTI data with these intents. For example, to extract the right cortex data from a `"xifti"` object named `xii` using the `ciftiTools` package, the code is

`xii$data$cortex_right`,

whereas the corresponding code for a `"cifti"` object named `cii` using the `cifti` package, shown below, is more verbose. This is because the `"cifti"` object more closely resembles the original structure of the CIFTI XML, which must be generalizable to all CIFTI intents.

```
all_bs <- sapply(cii$BrainModel, attr, "BrainStructure")
bm <- cii$BrainModel[[which(all_bs == "CIFTI_STRUCTURE_CORTEX_RIGHT")]]
cii$data[attr(bm, "IndexOffset") + seq(attr(bm, "IndexCount")),].
```

Another important difference is that `ciftiTools` integrates the Connectome Workbench and support for GIFTI surface geometry, which opens up more functionalities, particularly those listed in Panels B and C of **Figure 1**. In contrast, the `cifti` package does not have any external dependencies.

In summary, whereas `ciftiTools` makes it easy to work with dtseries, dscalar, or dlabel CIFTI files, users interested in working with other intent types, who want to read in the data structured similarly to the original CIFTI file, or who do not want to depend on the Connectome Workbench should look to `cifti` as an alternative. The R package `xml2` may also be helpful for parsing the XML metadata in CIFTI files at a low level (Wickham, Hester, and Ooms 2020).

### 4.3.2 The gifti R package

The `gifti` R package reads and writes GIFTI files, but does not provide further support (Muschelli 2020). `ciftiTools` depends on `gifti` for reading and writing GIFTI files, for example with `read_xifti2` and `write_xifti2`, and therefore builds on top of it to provide additional support for surface-based analyses including visualization and data manipulation. The `"xifti"` object also allows users to work with the corresponding left and right hemisphere data together, rather than having to separately process the left and right GIFTI files.

### 4.3.3 The freesurferformats and fsbrain R packages

The R package `freesurferformats` provides low-level reading and writing functionality for many different types of neuroimaging files, and `fsbrain` interfaces with it to offer higher-level reading and writing functions as well as a suite of visualization tools (Schäfer and Ecker 2021, 2020). The focus of these packages is on structural surface modalities. They are especially designed to support Freesurfer output, but are not limited to the FreeSurfer files or organization structures. There is limited support for CIFTI-format files. For example, the function for reading CIFTI files in `freesurferformats`, which is built on top of the `cifti` package, does not distinguish between the left and right hemispheres when reading in data from across the entire cortex. As with `cifti`, it cannot write out CIFTI files. Compared with `ciftiTools`, code to accomplish the same task can be more verbose. For example, to plot the cortical data with `ciftiTools` the command is

```
xii <- read_cifti(cii_fname, surfL_fname, surfR_fname)
plot(xii)
```

whereas for `freesurferformats` and `fsbrain` it is

```
data_l <- read.fs.morph.cifti(cii_fname, "lh")
data_r <- read.fs.morph.cifti(cii_fname, "rh")
surf_l <- read.fs.surface(surfL_fname)
surf_r <- read.fs.surface(surfR_fname)
mesh_l <- coloredmesh.from.preloaded.data(surf_l, data_l, hemi="lh")
mesh_r <- coloredmesh.from.preloaded.data(surf_r, data_r, hemi="rh")
brainviews("t4", list(mesh_l, mesh_r), draw_colorbar=TRUE)
```

### 4.3.4 The ggseg, ggseg3d, and ggsegExtra R packages

Brain atlases (i.e. segmentations or parcellations) can be plotted with 2-dimensional graphics using `ggseg` and 3-dimensional, interactive graphics using `ggseg3d` (Mowinckel and Vidal-Piñeiro 2020). `ggseg` relies on `ggplot2` whereas `ggseg3d` relies on `plotly`. Each package includes two parcellations—the Desikan-Killiany cortical atlas (Desikan et al. 2006) and the Fischl subcortical segmentation (Fischl et al. 2002)—and additional parcellations can be loaded with a separate package, `ggsegExtra`. While both `ciftiTools` and `ggseg3d` can display cortical data on a 3D mesh, `ggseg` and `ggseg3d` offer additional visualizations not provided by `ciftiTools`: `ggseg` can display cortical data as orthogonal slices, and `ggseg3d` can display subcortical data as 3D, interactive images. However, these visualizations are primarily useful for parcellated data. In contrast,

`ciftiTools` can display parcellated data but is not limited to it, having the ability to visualize the cortical surface and subcortical volume at the same resolution as the original data. Another important distinction is that `ciftiTools` is designed for CIFTI, GIFTI, and NIFTI file compatibility whereas `ggseg`, `ggseg3d`, and `ggsegExtra` are designed for Freesurfer-format file directories. In fact, to import new parcellations with `ggsegExtra`, an installation of Freesurfer is required. Conversion between Freesurfer-format file directories and CIFTI or GIFTI files is possible but non-trivial, so we recommend using `ggseg`, `ggseg3d`, and `ggsegExtra` to work with Freesurfer-format parcellations, and `ciftiTools` to work with CIFTI files, GIFTI files, and/or non-parcellated data.

## 4.4 Working with CIFTI files in other programming environments

Several tools have been developed in the Python (Van Rossum and Drake Jr 1995) and MATLAB (MATLAB 2021) programming languages to provide varying levels of support for CIFTI processing, analysis, and visualization. In this section we describe those listed in **Table 1** in greater detail.

### 4.4.1 Python

The `ciftify` module (Dickie et al. 2019) converts volumetric fMRI images with accompanying structural $T_1$-weighted files to the CIFTI format using `fmriprep` tools, allowing CIFTI-based Workbench and HCP tools to be used on data not originally in the CIFTI format. `ciftify` has functionality for analyzing and visualizing the resulting files, including a comparison with the original data for quality assurance. The fMRI data are not read in to the environment; rather, `ciftify` functions operate on the files directly and write any results or plots to files. A CIFTI file created by `ciftify` could be read in with `ciftiTools` for manipulation and interactive visualization.

The `NiBabel` library (Brett et al. 2016) supports reading, writing, and manipulation of many neuroimaging file formats, including CIFTI files of all intents. It includes methods for parsing the CIFTI files once read in. For example, the data from a specific brain structure can be extracted as a `numpy` array, the data matrix structure commonly used by machine learning tools in Python. It does not include support for visualizations. The `nilearn` toolbox (Huntenburg et al. 2017) does not currently support CIFTI files per se, but it has functions for projecting data onto surface geometry for visualization. Users can read in CIFTI and GIFTI data with `NiBabel` and then use `nilearn` to visualize it. `hcp-utils` (Janik 2020) provides high-level access to CIFTI files, including visualization, parcellation, and spatial masking. `hcp-utils` extends upon the functionality provided by `NiBabel` and `nilearn`, and thus requires both to be installed.

### 4.4.2 MATLAB

`cifti-matlab` is a software toolbox recommended by the HCP for working with CIFTI-format data within MATLAB (Oostenveld 2021). It does not depend on the Connectome Workbench and supports reading, writing, creation, and some manipulation of CIFTI files. It is currently under development. No visualization support for CIFTI-format data is currently available natively in MATLAB to our knowledge, though GIFTI-format data can be read, written, and visualized using the `GIFTI` library (Flandin 2021).

# 5 Discussion

Here we have presented and illustrated `ciftiTools`, an R package for reading, writing, visualizing, and manipulating CIFTI and GIFTI files. The stable version of `ciftiTools` is available via CRAN[1] as well as GitHub[2], where development and archived versions of the package, and a tutorial vignette, can also be found. The primary goal of `ciftiTools` is to provide a comprehensive, user-friendly programmatic and visualization tool supporting CIFTI- and GIFTI-based workflows. The neuroimaging community is currently in a period of active development of tools for working with grayordinates data. Today, a number of tools in MATLAB, Python, and R have been adapted to provide support for working with CIFTI files. However, `ciftiTools` is one of the few that has been developed specifically for CIFTI-format data. As such, the set of functionalities provided by `ciftiTools` for working with and visualizing CIFTI-format data is one of the most extensive and user-friendly among programmatic CIFTI-compatible tools. Moreover, the functionalities provided by `ciftiTools` will continue to expand as the needs of the user base evolve and become clear.

A secondary goal of `ciftiTools` is to facilitate the use of advanced statistical methods available in R for grayordinate-based fMRI analysis. Prior to the development of `ciftiTools`, CIFTI files could be read into R but not written out or directly visualized. R users were required to adopt a patchwork approach to visualization of analysis results, first writing results to a generic text or CSV file, reading that data into another program such as MATLAB or Python capable of writing out CIFTI-format data, then finally visualizing the results, e.g. via the Connectome Workbench. To state the obvious, this did not represent a convenient workflow for analysis of grayordinates data in R. By simplifying CIFTI- and GIFTI-based workflows, `ciftiTools` bridges a gap between existing statistical methods available in R and grayordinates-based MR data for which they are well suited to analyze. Bayesian techniques in particular hold promise for extracting more accurate, specific, and individualized insights by taking advantage of the improved alignment of structural and functional features across subjects, geodesic distances, and reduced dimensionality of grayordinate-based and surface-based MR data. As the statistics and neuroimaging communities working with R continue to grow, our hope is that `ciftiTools` will also serve as a foundation for the development and adoption of future Bayesian techniques and other advanced statistical methods for grayordinate-based and surface-based MR data.

# 6 Acknowledgements

This research was supported in part by NIH grant R01EB027119 from the National Institute of Biomedical Imaging and Bioengineering, and NIH grant 5R01NS060910-12 from the National Institute of Neurological Disorders and Stroke.

---

[1] https://cran.r-project.org/web/packages/ciftiTools/index.html
[2] https://github.com/mandymejia/ciftiTools/

# A  List of commonly used functions

Table 3 lists the most commonly-used functions in `ciftiTools`. The same list can be displayed in R Studio using the command `help("ciftiTools")`. Note that several of these functions also have aliases. For example, `readcii` and `read_cifti` are both equivalent to `read_xifti`, and `writecii` and `write_cifti` are both equivalent to `write_xifti`. These aliases are provided for convenience to users unfamiliar with the `"xifti"` nomenclature.

| **Reading in CIFTI or GIFTI data** | |
|---|---|
| read_xifti | Read in a CIFTI file as a "xifti" |
| read_xifti2 | Read in GIFTI files as a "xifti" |
| as.xifti | Combine numeric data to form a"xifti" |
| read_surf | Read in a surface GIFTI file as a "surf" |
| info_cifti | Read the metadata in a CIFTI file |
| load_surf | Read in a surface included in "ciftiTools" |
| load_parc | Read in a parcellation included in "ciftiTools" |
| **Writing CIFTI or GIFTI data** | |
| write_cifti | Write a "xifti" to a CIFTI file |
| write_cifti2 | Write a "xifti" to GIFTI and NIFTI files |
| write_metric_gifti | Write a numeric data matrix to a metric GIFTI file |
| write_surf_gifti | Write a "surf" to a surface GIFTI file |
| write_subcort_nifti | Write subcortical data to NIFTI files |
| separate_cifti | Separate a CIFTI file into GIFTI and NIFTI files |
| **Manipulating a "xifti"** | |
| apply_xifti | Apply a function along the rows or columns of the "xifti" data matrix |
| combine_xifti | Combine multiple "xifti"s with non-overlapping brain structures |
| convert_xifti | Convert the intent of a "xifti" |
| merge_xifti | Concatenate data matrices from multiple "xifti"s |
| newdata_xifti | Replace the data matrix in a "xifti" |
| remove_xifti | Remove a brain structure or surface from a "xifti" |
| select_xifti | Select data matrix columns of a "xifti" |
| transform_xifti | Apply a univariate transformation to a "xifti" or pair of "xifti"s |
| add_surf | Add surfaces to a "xifti" |
| move_from_mwall | Move medial wall vertices back into the "xifti" data matrix |
| move_to_mwall | Move rows with a certain value into the "xifti" medial wall mask |
| **S3 methods for "xifti"s** | |
| summary and print | Summarize the contents |
| as.matrix | Convert data to a locations by measurements numeric matrix |
| dim | Obtain number of locations and number of measurements |
| plot | Visualize the cortical surface and/or subcortical data |
| +, -, *, /, ^, %%, %/%, ==, != | Operation between a "xifti" and a scalar, or between two "xifti"s |
| abs, ceiling, exp, floor, log, round, sign, sqrt | Univariate transformation of "xifti" data |
| **Working with "surf"s** | |
| read_surf | Read in a surface GIFTI file as a "surf" |
| is.surf | Verify a "surf" |
| write_surf_gifti | Write a "surf" to a surface GIFTI file |
| view_surf | Visualize a "surf" |
| resample_surf | Resample a "surf" |
| rotate_surf | Rotate the geometry of a "surf" |

**Table 3:** Commonly used functions currently available in `ciftiTools`.

# B  The medial wall in CIFTI files, GIFTI files, and "xifti" objects

For both the left and right cortical surfaces, CIFTI files have a metadata entry named "VertexIndices" which lists, in order, the index of each vertex represented by the data matrix. These vertices form the "region of interest" (ROI). Any vertex not in "VertexIndices" does not have data for it. Typically these omitted vertices form the medial wall, for which no gray matter data is available. However, if the CIFTI file only represents a certain portion of the brain surface, it will have a smaller ROI and therefore more vertices omitted from "VertexIndices." Another metadata entry named "SurfaceNumberOfVertices" gives the total number of vertices on the associated surface geometry, so that no matter the size of the ROI, the resolution of the data can be known.

When a CIFTI file with surface data is separated into GIFTI files with the `ciftiTools` function `separate_cifti`, each hemisphere yields two GIFTI files: one for the data, and one for the ROI mask. For the left cortex, for example, these are named `"cortexL"` and `"ROIcortexL"` in the vector of file names returned by `separate_cifti`. The data GIFTI will be padded with zeroes in place of the omitted vertices, and the ROI GIFTI will have ones for vertices inside the ROI, and zeroes for those outside it. Therefore, the ROI GIFTI is necessary to distinguish which zeroes in the data GIFTI are actual zero-valued vertices, and which represent excluded vertices.

The `"xifti"` object nomenclature assumes that any vertex outside the ROI is part of the medial wall, since this is the case in most applications of CIFTI files. For a `"xifti"` named `xii` with left cortex data, `xii$meta$cortex$medial_wall_mask$left` will be a `TRUE`/`FALSE` vector the same length as the associated surface geometry, with `TRUE` values indicating vertices within the ROI. This is similar to the ROI GIFTI. When the mask is not present, it is assumed to be `TRUE` for all vertices. The data matrix in `xii` will not use padding zeroes to represent the vertices outside the ROI, since that would be redundant with the medial wall mask. In this sense, the data matrix is similar to that of CIFTI files. Note this means that two CIFTIs with the same resolution but different medial wall masks will have data matrices with different numbers of rows.

To clarify with an example: if a CIFTI file has $m$ measurements of $v_0$ data locations across the left cortex, and if the surface geometry of the left cortex has $v$ total vertices, then its data matrix will be $v_0 \times m$. "VertexIndices" will list $v_0$ indices, and "SurfaceNumberOfVertices" will be $v$. When this CIFTI file is exported to GIFTI files, the data GIFTI will have $m$ measurements of $v$-length vectors padded with zeroes, and the ROI GIFTI will have one $v$-length measurement indicating which vertices in the data GIFTI are inside the ROI. When this CIFTI file is read in with `xii <- read_xifti(...)`, `dim(xii)` will be equal to `c(v_0, m)` and `xii$meta$cortex$medial_wall_mask$left` will be a $v$-length logical vector with $v_0$ `TRUE` values.

Sometimes it is useful to pad the data matrix with `NA`, `0`, or some other value, instead of using the medial wall mask in the metadata. In this case, the `ciftiTools` function `move_from_mwall` can delete the medial wall mask and add rows to the data matrix to pad it. Conversely, `move_to_mwall` can move vertices with a certain value from the data matrix into the medial wall mask.

# C  Embedding ciftiTools plots in Markdown documents

R Markdown and the `knitr` package are used to create linear, reproducible reports by allowing authors to weave together a mix of Markdown, LaTeX, HTML, and R code (Allaire et al. 2021; Xie 2021). `ciftiTools` plots can be embedded in R Markdown documents in a number of ways. As demonstrated in Section 3, `view_xifti_surface` and `view_xifti_volume` can both write PNG files which may be embedded in R Markdown documents like any other image file. In the example we used the ciftiTools function `view_comp`. Other options are `knitr::include_graphics("plot.png")` or `graphics::rasterImage(png::readPNG("plot.png"))` inside a code chunk, and `` outside a code chunk. If the only purpose of saving a PNG file is to embed it in an R Markdown document, the user may write the PNG to a temporary directory or file with `tempdir()` or `tempfile()`. For methods that work inside a code chunk, the resulting image size can be controlled with chunk options, including `fig.width` and `fig.height`.

`view_xifti_surface` has additional options that work when the resulting document is an HTML file. First, snapshots of the OpenGL window can be directly embedded. This allows plots to be rendered without needing to save an intermediate file. Embedding snapshots requires that a setup function be executed:

```
library(rgl)
rgl::setupKnitr()
rgl::rgl.open(); rgl::rgl.close() # first window in doc. may not render properly
```

after which a code chunk can produce a cortical surface plot using the chunk options `rgl=TRUE` and `format="png"`:

````
```{r, fig.cap="Snapshot", rgl=TRUE, format="png", eval=FALSE} plot(xii) ```
````

The second option from `view_xifti_surface` for interactive plots is the HTML widget. Unlike the OpenGL snapshot, an HTML widget will remain interactive when embedded. No setup is required; a chunk just needs to have a call to `view_xifti_surface` which produces a widget (see Section 2.3), such as in the following chunk.

````
```{r, eval=FALSE} plot(xii, idx=seq(3)) ```
````

These `ciftiTools` features make it easier for users to write reproducible, interactive documents for the analysis and visualization of grayordinates data. The numerous math and manipulation functions implemented in `ciftiTools` also enable users to perform many common operations in single lines of code, allowing scripts in R Markdown reports to be more concise and easy for readers to follow.

# D  Data included in ciftiTools

Several external datasets are included within the `ciftiTools` package:

- Example CIFTI files from NITRC are used for demonstration and codebase testing.[3] These are accessed with the command `ciftiTools.files()$cifti`.
- Surface geometry GIFTI files from the HCP are used for visualizing and smoothing cortical data, should the user not provide their own. Three different surfaces are available–"inflated," "very inflated," and "midthickness"–and each has about 32,000 vertices per hemisphere. These are provided according to the HCP Data Use Terms: "Data were provided [in part] by the Human Connectome Project, WU-Minn Consortium (Principal Investigators: David Van Essen and Kamil Ugurbil; 1U54MH091657) funded by the 16 NIH Institutes and Centers that support the NIH Blueprint for Neuroscience Research; and by the McDonnell Center for Systems Neuroscience at Washington University" (Glasser et al. 2013).[4] The files are accessed with the command `ciftiTools.files()$surf` or `load_surf`.
- Several cortical parcellations are included for convenience: the Schaefer parcellation at 100, 400, or 1000 parcels resolutions (Schaefer et al. 2018), and the Yeo parcellation at 7 networks (51 components) or 17 networks (114 components) resolutions (Thomas Yeo et al. 2011). These can be loaded with the function `load_parc`.
- The $T_1$ image of the MNI ICBM 152 non-linear 6th Generation Symmetric Average Brain Stereotaxic Registration Model is used in the subcortical visualizations as a spatial reference upon which the data is overlaid (Grabner et al. 2006).
- FreeSurfer color assignments for subcortical structures are used when writing CIFTIs files with subcortical data.[5]

---

[3] https://www.nitrc.org/frs/?group_id=454
[4] https://balsa.wustl.edu/reference/show/pkXDZ
[5] https://surfer.nmr.mgh.harvard.edu/fswiki/FsTutorial/AnatomicalROI/FreeSurferColorLUT

# E  Cortical surface visualization comparison

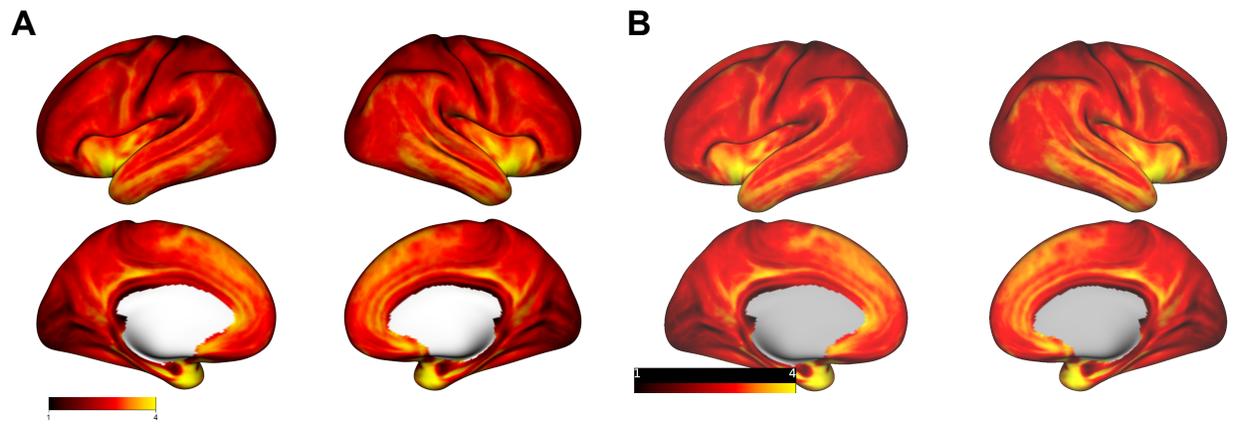

**Figure 6: Comparison of `ciftiTools` and Connectome Workbench cortical surface visualizations.** The same data and surface are plotted with `ciftiTools` (A) and the `wb_view` Connectome Workbench GUI (B). The same color palettes and limits were used, and the background of the Connectome Workbench plot was set to white to match the default for `ciftiTools`. The plot dimensions and surface sizes differ only slightly, and these parameters can be adjusted with both tools. Otherwise, the main differences are that (1) the color bar has a different appearance and is lower in the `ciftiTools` plot, such that it does not overlap with the left medial view, (2) missing values are white by default in `ciftiTools`, but light gray for the Connectome Workbench, and (3) there is more contrast–brighter highlights and darker shadows–in the 3-dimensional shading in the `ciftiTools` plot.

**Figure 6** compares the cortex surface data plots for `ciftiTools` and the Connectome Workbench GUI, `wb_view`.

# Expanded demonstration: Computing parcellated seed correlations

```r
library(rgl)
rgl::setupKnitr()

# Sometimes the first OpenGL window does not render properly.
rgl::rgl.open(); rgl::rgl.close()
```

In this demonstration we will use `ciftiTools` to compute seed correlations for a resting-state fMRI scan from the Human Connectome Project (HCP) (Van Essen et al. 2013). Our seed will be the posterior cingulate cortex (PCC), a highly-connected node of the default mode network (DMN). We will use functional parcellations created by Schaefer et al. (2018) to define the PCC and other brain regions. The analysis is divided into five steps: (1) reading in and smoothing the fMRI data; (2) reading in the parcellation; (3) cleaning the fMRI data; (4) calculating seed correlations; and (5) visualizing and saving the results.

This document expands on the abridged demonstration by adding a data cleaning section, including the subcortical data, and using more functions from `ciftiTools`. Since the data are being cleaned, we use the minimally pre-processed (MPP) rfMRI data instead of the data pre-processed with ICA-FIX (Salimi-Khorshidi et al. 2014) which was used in the abridged demonstration. We also knit this document to an HTML file instead of a PDF, which allows for the integration of interactive plots.

# 1 Reading in and smoothing the fMRI data

We begin by loading `ciftiTools` and pointing to the Connectome Workbench folder.

```r
library(ciftiTools)
ciftiTools.setOption("wb_path", "workbench")
```

```
## Using this Workbench path: 'workbench/bin_windows64/wb_command.exe'.
```

Now we can read in the CIFTI file containing the resting-state fMRI (rfMRI) data with `read_xifti`. By default, only the left and right cortex are read in, so we set `brainstructures="all"` to include the subcortex too.

```r
xii <- read_xifti("Data/rfMRI.dtseries.nii", brainstructures="all")
```

`xii` stores the CIFTI data as a `"xifti"` object. `summary`, or equivalently invoking the implicit `print` S3 method, will produce an overview of its contents:

```r
xii
```

```
## ====CIFTI METADATA===================
## Intent:           3002 (dtseries)
## - time step       0.72 (seconds)
## - time start      0
## Measurements:     1200 columns
```

```
## 
## ====BRAIN STRUCTURES=================
## - left cortex     29696 data vertices
##                   2796 medial wall vertices (32492 total)
## 
## - right cortex    29716 data vertices
##                   2776 medial wall vertices (32492 total)
## 
## - subcortex       31870 data voxels
##                   subcortical structures and number of voxels in each:
##                     Cortex-L (0), Cortex-R (0),
##                     Accumbens-L (135), Accumbens-R (140),
##                     Amygdala-L (315), Amygdala-R (332),
##                     Brain Stem (3472),
##                     Caudate-L (728), Caudate-R (755),
##                     Cerebellum-L (8709), Cerebellum-R (9144),
##                     Diencephalon-L (706), Diencephalon-R (712),
##                     Hippocampus-L (764), Hippocampus-R (795),
##                     Pallidum-L (297), Pallidum-R (260),
##                     Putamen-L (1060), Putamen-R (1010),
##                     Thalamus-L (1288), Thalamus-R (1248).
```

The S3 method `as.matrix` will return the timeseries data in matrix form, with locations (vertices and/or voxels) along the rows and measurements (timepoints) along the columns. Functions like `dim` which coerce their inputs to data matrices will behave accordingly:

```
dim(xii)
```

```
## [1] 91282  1200
```

We can also view and edit the `"xifti"` directly. It is a nested list:

```
str(xii, nchar.max = 32)
```

```
## List of 3
##  $ data:List of 3
##   ..$ cortex_left : num [1:29696, 1:1200] 9720 10430 8868 14286 10128 ...
##   ..$ cortex_right: num [1:29716, 1:1200] 11520 8820 8702 13177 11471 ...
##   ..$ subcort     : num [1:31870, 1:1200] 11835 11835 11835 11835 11835 ...
##  $ surf:List of 2
##   ..$ cortex_left : NULL
##   ..$ cortex_right: NULL
##  $ meta:List of 3
##   ..$ cortex :List of 2
##   .. ..$ medial_wall_mask:List of 2
##   .. .. ..$ left : logi [1:32492] TRUE TRUE TRUE TRUE TRUE TRUE ...
##   .. .. ..$ right: logi [1:32492] TRUE TRUE TRUE TRUE TRUE TRUE ...
##   .. ..$ resamp_res      : NULL
##   ..$ subcort:List of 4
##   .. ..$ labels    : Factor w/ 21 levels "Cortex-L","Cortex-R",..: 7 7 7 7 7 7 7 7 7 7 ...
##   .. ..$ mask      : logi [1:91, 1:109, 1:91] FALSE FALSE FALSE| __truncated__ ...
##   .. ..$ trans_mat : num [1:4, 1:4] -2 0 0 0 0 2 0 0 0 0 ...
```

```
##     .. ..$ trans_units: chr "mm"
##     ..$ cifti    :List of 6
##     .. ..$ intent          : num 3002
##     .. ..$ brainstructures: chr [1:3] "left" "right" "subcortical"
##     .. ..$ time_start      : num 0
##     .. ..$ time_step       : num 0.72
##     .. ..$ time_unit       : chr "second"
##     .. ..$ misc            :List of 4
##     .. .. ..$ ParentProvenance : chr "/HCP/hcpdb/buil"| __truncated__
##     .. .. ..$ ProgramProvenance: chr "Workbench\nVers"| __truncated__
##     .. .. ..$ Provenance       : chr "/NRG/BlueArc/nr"| __truncated__
##     .. .. ..$ WorkingDirectory : chr "/HCP/hcpdb/buil"| __truncated__
##  - attr(*, "class")= chr "xifti"
```

For example, let's add an entry to the miscellaneous metadata:

```
xii$meta$cifti$misc$ScanLocation <- "Candy Land"
str(xii$meta$cifti$misc, nchar.max = 32)
```

```
## List of 5
##  $ ParentProvenance : chr "/HCP/hcpdb/buil"| __truncated__
##  $ ProgramProvenance: chr "Workbench\nVers"| __truncated__
##  $ Provenance       : chr "/NRG/BlueArc/nr"| __truncated__
##  $ WorkingDirectory : chr "/HCP/hcpdb/buil"| __truncated__
##  $ ScanLocation     : chr "Candy Land"
```

Now we will spatially smooth the data with `smooth_xifti` using a geodesic Gaussian smoother. We will use the default 5 mm FWHM width, but other levels of smoothing are possible using the `surf_FWHM` argument. Note we smooth the original CIFTI file and then read it in, rather than smoothing `xii` which would be slower.

```
smooth_xifti("Data/rfMRI.dtseries.nii", "Data/rfMRI_sm.dtseries.nii")
xii_sm <- read_xifti("Data/rfMRI_sm.dtseries.nii", brainstructures="all")
```

We can plot the first columns of `xii` and `xii_sm` to get an idea of the data scale, range, and effect of smoothing. `plot` will create a 3D visualization of the cortical data by invoking `view_xifti_surface`. There are several options for displaying the plots. Here we will plot each `"xifti"` separately, save the two plots to PNG files, and then composite the PNG files together with `view_comp`.

```
plot(xii, title="Original", fname="PlotsEx/rfMRI.png")
```

```
## `zlim` not provided: using color range 0 - 14200 (data limits: 740 - 21100).
```

```
plot(xii_sm, title="Smoothed", fname="PlotsEx/rfMRI_sm.png")
```

```
## `zlim` not provided: using color range 0 - 13900 (data limits: 2060 - 16900).
```

```
view_comp(
  c("PlotsEx/rfMRI.png", "PlotsEx/rfMRI_sub.png",
    "PlotsEx/rfMRI_sm.png", "PlotsEx/rfMRI_sm_sub.png"),
```

```
"pair"
)
```

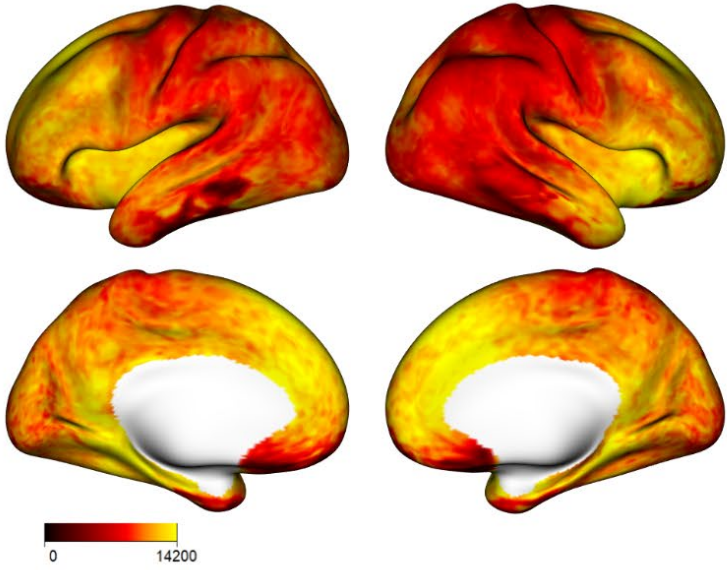

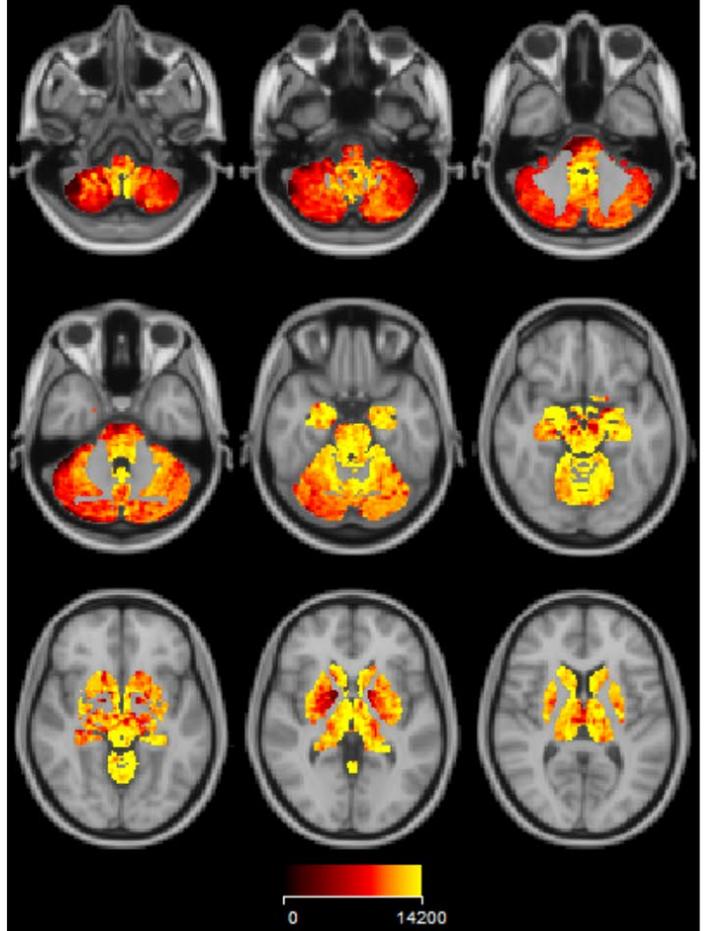

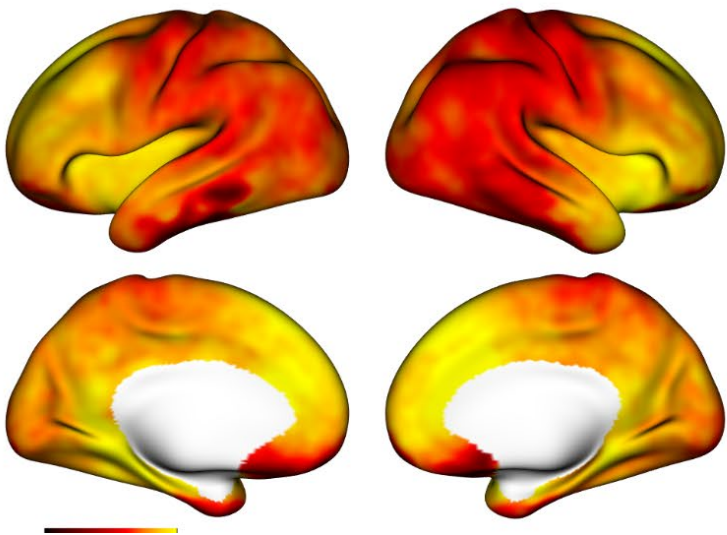

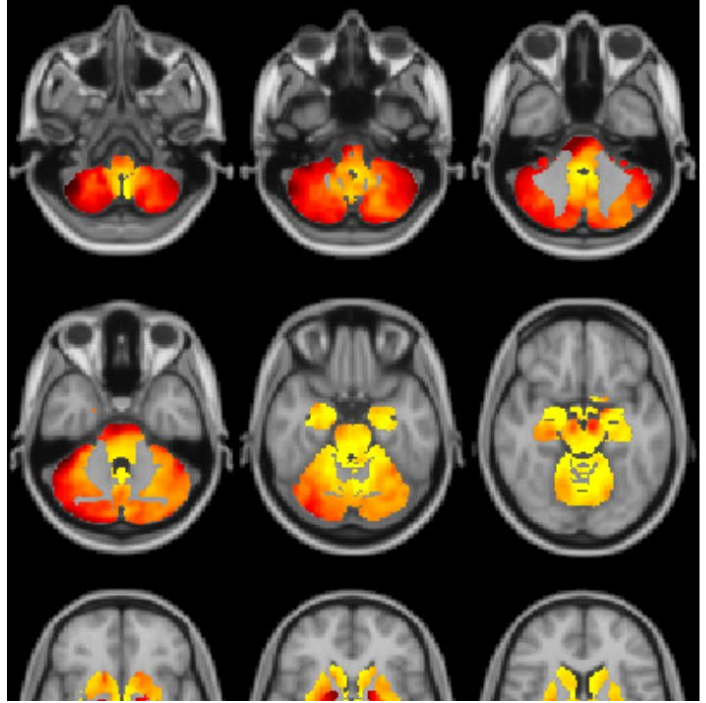

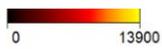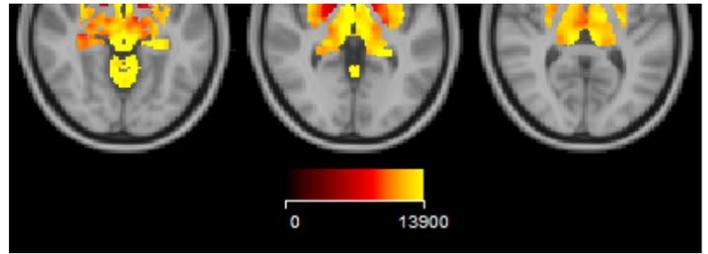

Figure 1.1: First fMRI volume before and after smoothing with smooth_xifti

The color scale limits have been automatically determined based on the data, so they are close but slightly different. Nonetheless, the data on the right are clearly similar to the original except smoother across space.

We can also embed the plots one at a time directly, without saving to a file. (This is true for both HTML and PDF documents for the subcortex plot, but only true for HTML documents for the cortex plot.) We will demonstrate this below while also showing the use of different color palettes:

```
view_xifti_surface(xii, title="Smoothed", colors="viridis")
```

```
## `zlim` not provided: using color range 0 - 13600 (data limits: 1000 - 17700).
```

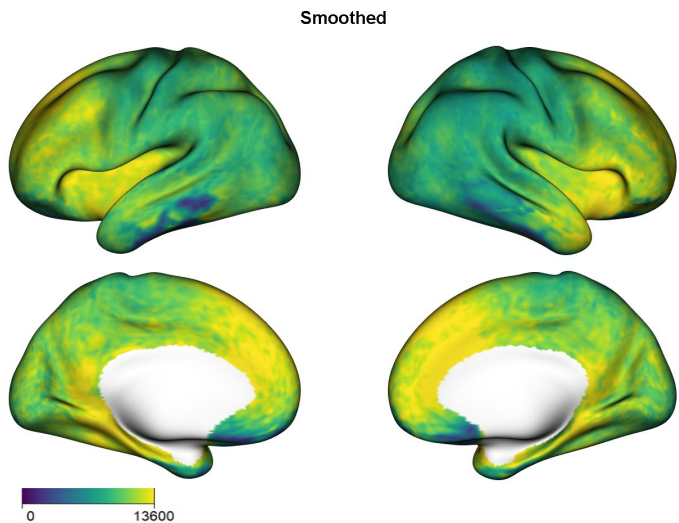

Figure 1.2: First fMRI volume after smoothing with smooth_xifti (cortex and viridis palette)

```
view_xifti_volume(xii, title="Smoothed", colors="BuPu")
```

```
## `zlim` not provided: using color range 0 - 15000 (data limits: 740 - 21100).
```

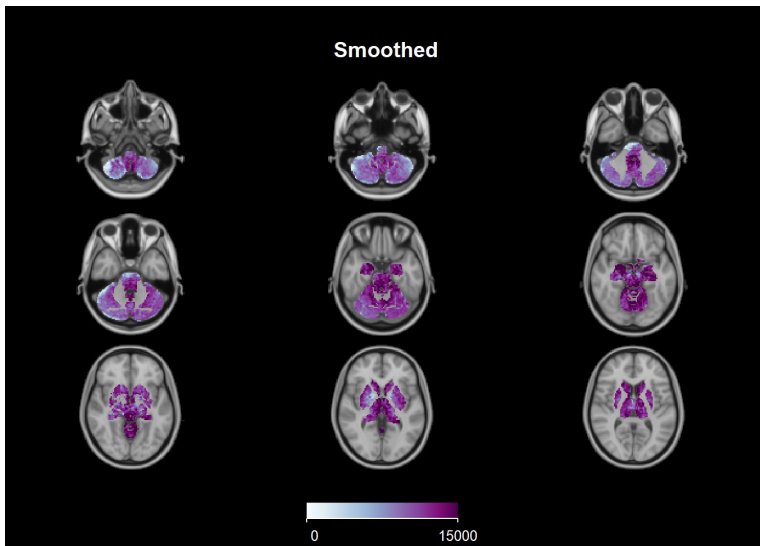

Figure 1.3: First fMRI volume after smoothing with smooth_xifti (subcortex and RColorBrewer blue-purple palette)

# 2 Reading in the parcellation

Now we will load the parcellation. For convenience, ciftiTools includes a few commonly used parcellations, which can be read in using `load_parc` (Schaefer et al. 2018; Thomas Yeo et al. 2011). Here we will use the Schaefer 400 parcellation. Other parcellations in CIFTI or GIFTI format can be read in using `read_xifti` or `read_xifti2`, respectively.

```
parc <- load_parc("Schaefer_400")
```

Again, we can get an overview of the `"xifti"` object `parc` by using `summary`.

```
parc
```

```
## ====CIFTI METADATA===================
## Intent:          3007 (dlabel)
## - names          "parcels"
## Measurements:    1 column
## 
## ====BRAIN STRUCTURES=================
## - left cortex    32492 data vertices
## 
## - right cortex   32492 data vertices
```

The parcellation includes data for the left and right cortical hemispheres at the same resolution of the rfMRI data (approx. 32,000 vertices per hemisphere). The parcellation has the "dlabel" intent which indicates that it contains label i.e. categorical data. Specifically, for each measurement (column) every data location is assigned an integer "key" which corresponds to a certain label and color. A table in the metadata gives the label and color of each key, for each measurement.

This file has only one measurement, `"parcels"`. Its keys range from 0 to 400, with 0 indicating the medial wall and 1-400 indicating a specific parcel. When visualizing the parcellation with `plot`, each parcel will be colored according to the label table colors by default. In this parcellation, parcels are grouped by network, and those in the same network are colored identically. Therefore, adjacent parcels in the same network are impossible to delineate by color. We can use the `border` argument to draw outlines between vertices of different keys, allowing us to visualize both the networks, according to color, and the individual parcels, according to outlines.

We will also separately visualize parcel 14, which corresponds closely to the PCC. We use `transform_xifti` to create a mask of only that parcel, and `merge_xifti` to create a new `"xifti"` with two measurements: the original parcellation and the masked PCC. Since this is an HTML document, we can then embed an interactive HTML widget in which the two measurements can be toggled between (). We will resample the data before plotting to reduce the file size of the HTML document, since the widget will be large otherwise.

```r
parc_PCC <- transform_xifti(parc, function(x){ ifelse(x == 14, 14, 0) })
parc2 <- merge_xifti(parc, parc_PCC)
parc2 <- resample_xifti(parc2, resamp_res=12000)
```

```
## Separating CIFTI file.
## Time difference of 6.689175 secs
## Resampling CIFTI file.
## Time difference of 6.131946 secs
## Merging components into a CIFTI file...
## Time difference of 0.4521921 secs
```

```r
plot(parc2, idx=seq(2), title = c("Schaefer 400", "PCC"), borders="black")
```

```
## Color labels from first requested column will be used.
```

Recall that the fMRI data include the subcortex, whereas the Schaefer parcellation does not. Instead of ignoring the subcortical data, we will treat each of the 19 subcortical structures as a brain region to bring the total number of seed correlations to 419.

# 3 Cleaning the fMRI data

We now perform a couple data cleaning steps to obtain more accurate seed correlation estimates. First, we detrend the data using nuisance regression with discrete cosine transform (DCT) bases to remove low-frequency variations which do not reflect neuronal signal. If ignored, similar drifts in different regions of the brain can lead to inflated estimates of seed correlation. We use the `fMRIscrub` functions `dct_convert` to determine the number of DCT bases that approximates a highpass filter of 0.01 Hz, `dct_bases` to obtain DCT bases with the same length as our data, and `nuisance_regression` to perform the nuisance regression. We also use the `ciftiTools` function `newdata_xifti` to format the result of `nuisance_regression` back into a `"xifti"` object.

```
library(fMRIscrub)

n_time <- ncol(xii_sm)
n_dct <- fMRIscrub::dct_convert(n_time, TR=.72, f=.008)
nreg <- cbind(1, fMRIscrub::dct_bases(n_time, n_dct))
mclean <- fMRIscrub::nuisance_regression(as.matrix(xii_sm), nreg)
xii_clean <- newdata_xifti(xii_sm, mclean)
```

We visualize the effect of detrending by looking at the timecourses for two different vertices before and after detrending .

```
par(mfrow = c(2, 2))
lplot <- function(x, ...){ plot(x, type="l",lty=1,ylab="Value",xlab="Time", ...) }

lplot(as.matrix(xii_sm)[1000,], main="1,000th vertex: before detrending")
lplot(as.matrix(xii_clean)[1000,], main="1,000th vertex: after detrending")

lplot(as.matrix(xii_sm)[20000,], main="20,000th vertex: before detrending")
lplot(as.matrix(xii_clean)[20000,], main="20,000th vertex: after detrending")
```

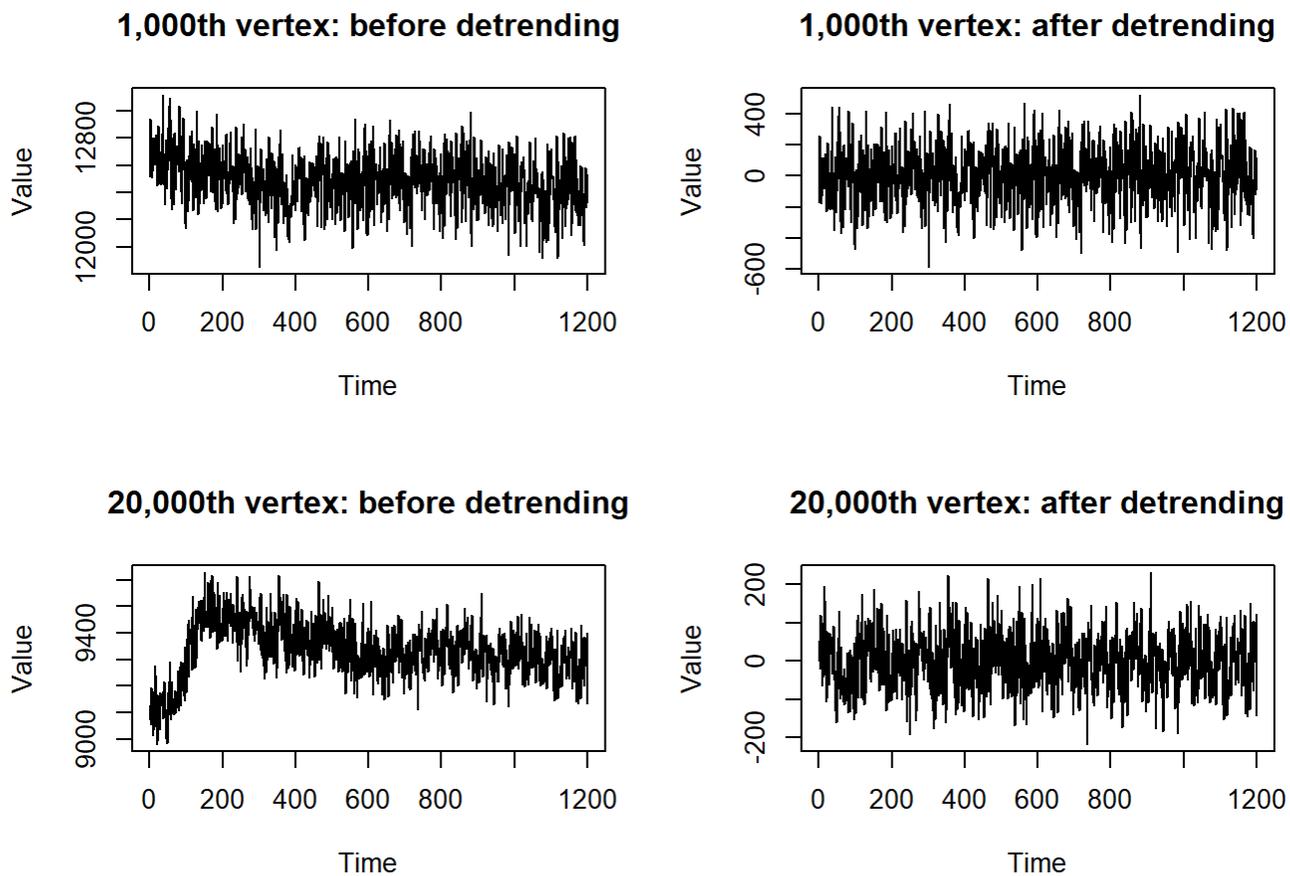

Figure 3.1: Two vertices before and after detrending

Next, we scrub volumes suspected of artifact contamination by thresholding DVARS (Afyouni and Nichols 2018), a measure of signal intensity change across the brain. Head motion and other transient artifact sources tend to induce variations of much greater magnitude than neuronal signals. Thus, by removing volumes with high DVARS, we ignore volumes likely contaminated by artifacts, which can otherwise influence seed correlations in unpredictable ways. In the following code, we compute DVARS using the `DVARS` function from `fMRIscrub`, and plot it using the S3 `plot` method for the result .

```
dv <- fMRIscrub::DVARS(xii_clean)
plot(dv, height=2)
```

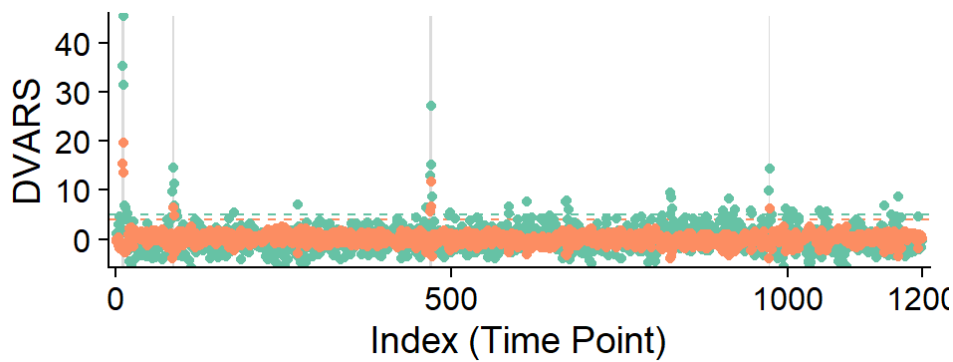

Figure 3.2: DVARS timecourse for the fMRI run. Orange and green dots represent the two types of DVARS measures. Volumes exceeding thresholds for both measures are indicated by the vertical grey highlights.

We remove the flagged volumes using `select_xifti`. We can count how many remain with `dim`: 1191, so 9 volumes were scrubbed.

```
outs <- dv$outlier_flag$Dual
xii_clean <- select_xifti(xii_clean, which(!outs))
dim(xii_clean)
```

```
## [1] 91282  1191
```

# 4 Computing seed correlations

Now that we have the cleaned rfMRI data and parcellation, we can compute the seed correlations. We begin by obtaining the mean timeseries of each brain region (400 Schaefer cortical parcels and 19 subcortical structures). This is easier done with the parcel data directly, rather than with the `parc` `"xifti"` object, so we will start by converting the parcellation to a vector. Then, we'll replace the medial wall vertices in `xii_clean` with `NA` data values so that the length of the parcellation vector, `parc_vec`, will match the number of cortical data vertices in `xii_clean`.

```
parc_vec <- c(as.matrix(parc))
length(parc_vec)
```

```
## [1] 64984
```

```
xii_clean <- move_from_mwall(xii_clean, NA)
dim(xii_clean)
```

```
## [1] 96854  1191
```

We will append the vector `sub_keys` to `parc_vec` in order to account for the subcortical voxels in the rfMRI data. The values of `sub_keys` will be between 401 and 419, each of which correspond to a subcortical structure. Note we subtract 2 because the first two labels are empty.

```
sub_keys <- as.numeric(xii_clean$meta$subcort$labels) - 2
sub_keys <- 400 + sub_keys
brain_vec <- c(parc_vec, sub_keys)
```

Now for each brain region we will obtain the mean time series across its constituent vertices or voxels. Specifically, we will pre-allocate a matrix `xii_pmean`, and then for each brain region we will obtain the fMRI timeseries for all its vertices or voxels, take the average across space while ignoring medial wall values, and then save the resulting timeseries to a row of `xii_pmean`.

```
xii_mat <- as.matrix(xii_clean)
xii_pmean <- matrix(nrow=400 + 19, ncol=ncol(xii_mat))
for (p in 1:(400 + 19)) {
  data_p <- xii_mat[brain_vec==p,]
  xii_pmean[p, ] <- colMeans(data_p, na.rm=TRUE)
}
```

We then calculate the seed correlations between the PCC parcel and every other brain region.

```
seed_cor <- cor(t(xii_pmean))[,14]
```

# 5 Visualizing and saving the results

We can visualize the results by creating a new `"xifti"` with `newdata_xifti`. It will contain one measurement (column) in which the value of each vertex will be the seed correlation for the brain region it belongs to. Notice how in the previous subsection we converted the parcellation `"xifti"` to numeric data with `as.matrix`, and after performing several operations are now returning to a `"xifti"` object with `newdata_xifti`. This is a workflow to use when it is not possible to obtain the result with only the operations built in to `ciftiTools`.

First we'll convert `parc_vec`, which gives the parcel key of each vertex, to `xii_seed`, which gives the seed correlation of the parcel each vertex belongs to. Then we will initialize a one-column `"xifti"` using `select_xifti` and replace its data with `xii_seed` using `newdata_xifti`.

```
xii_seed <- c(NA, seed_cor)[brain_vec+1]
xii1 <- select_xifti(xii_clean * 0, 1)
xii_seed <- newdata_xifti(xii1, xii_seed)
```

Now we can use the `"xifti"` S3 method `plot` again. We'll use it twice to demonstrate the ability of `view_xifti` to automatically choose an appropriate `zlim`.

```
plot(xii_seed, zlim=c(-1, 1), title="Set zlim", fname="PlotsEx/seed.png")
plot(xii_seed, title="Auto zlim", fname="PlotsEx/seed2.png")
```

```
## `zlim` not provided: using color range -0.571 - 0.571 (data limits: -0.244 - 1).
```

```
view_comp(
  c("PlotsEx/seed.png", "PlotsEx/seed_sub.png",
    "PlotsEx/seed2.png", "PlotsEx/seed2_sub.png"),
  "pair"
)
```

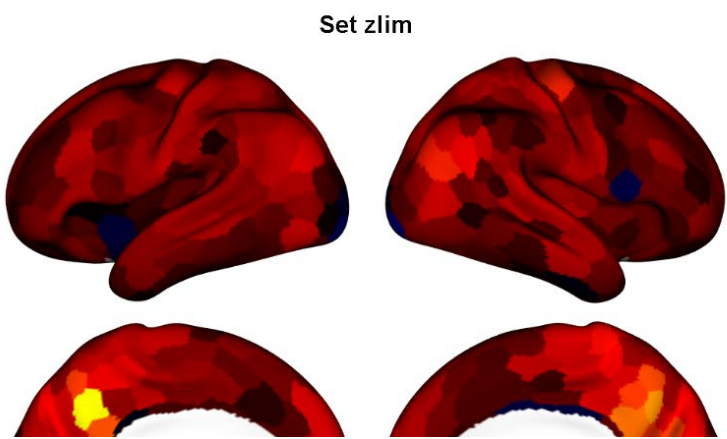
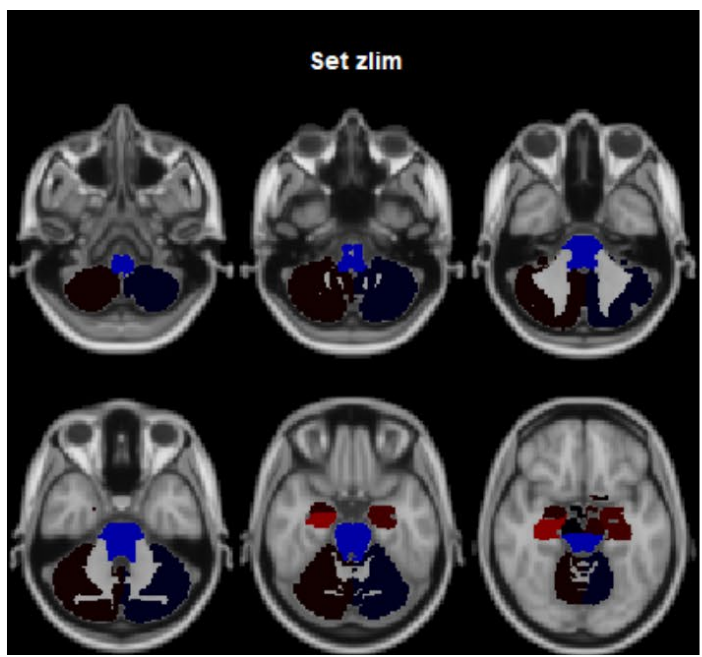

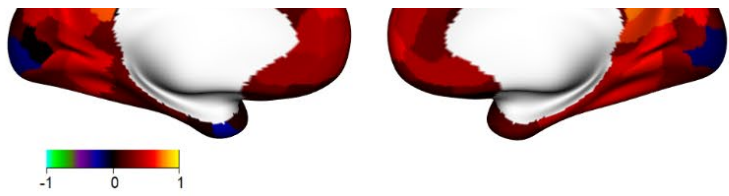
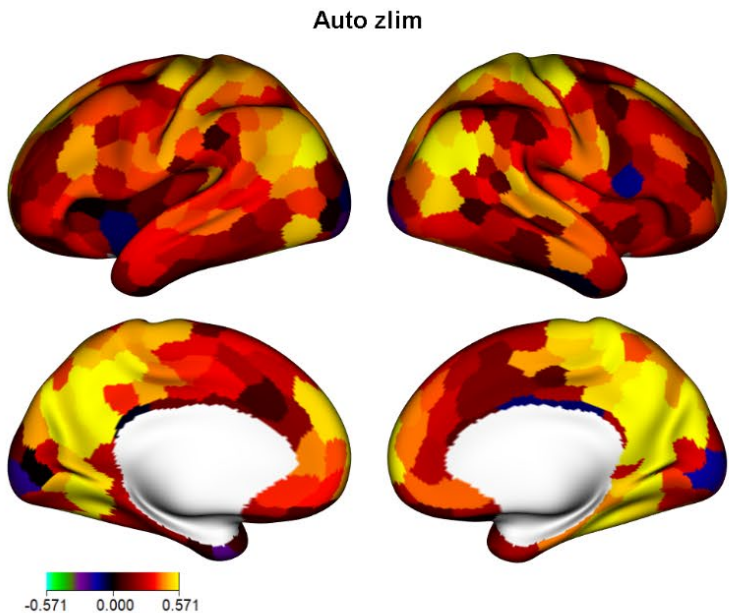
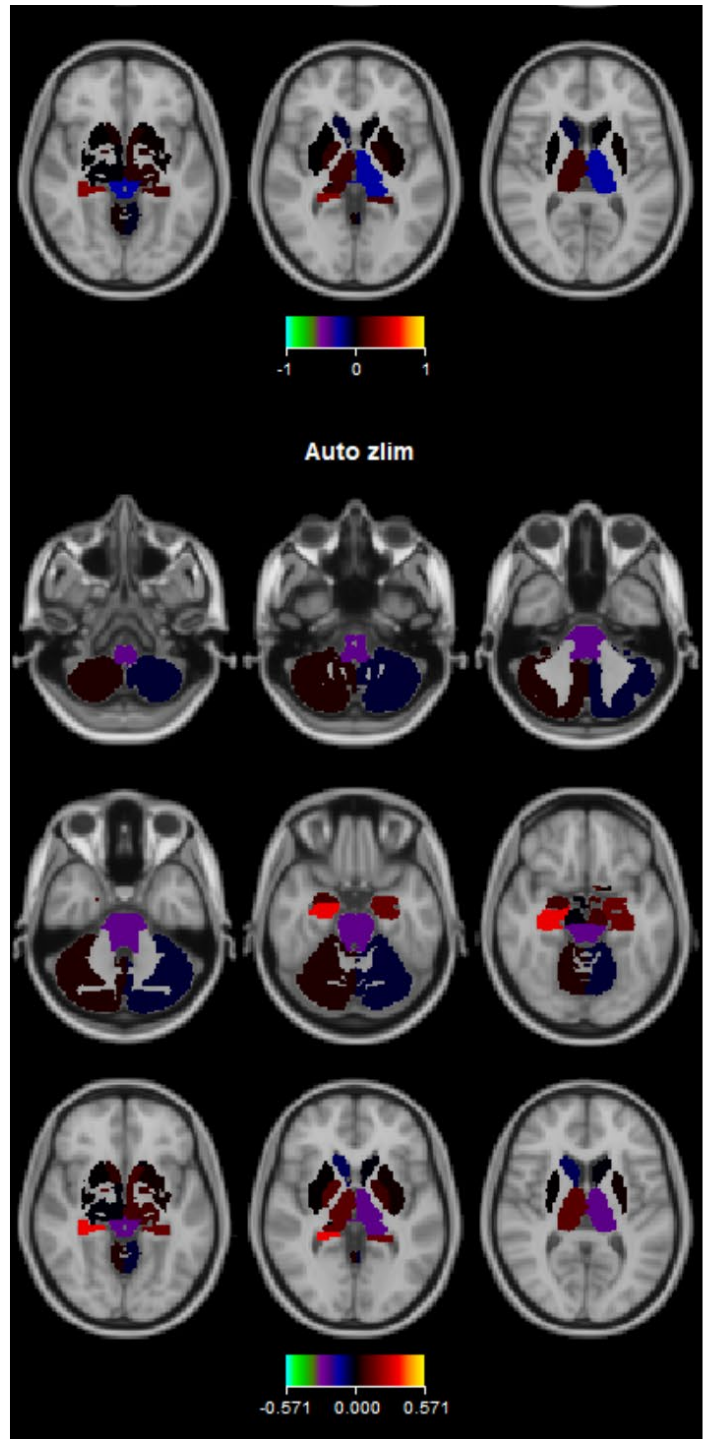

Figure 5.1: PCC seed correlation

As expected, we find that most seed correlation values are positive, and the brain regions with the highest seed correlation are components of the DMN. The seed correlation values are shifted positively compared to those from the abridged analysis in the paper, a difference that can be attributed to the different pre-processing: ICA-FIX in the abridged analysis and DCT detrending and scrubbing in this expanded analysis.

Lastly, we can save the result of our analysis with `write_xifti`. We will write the data to a "dscalar" CIFTI because it does not represent a timeseries ("dtseries") or categorical/label data ("dlabel").

```
write_xifti(xii_seed, "Data/PCC_seedCorEx.dscalar.nii")
```

```
## Writing left cortex.
```

```
## Writing right cortex.
## Writing subcortical data and labels.
## Creating CIFTI file from separated components.
```

```
file.exists("Data/PCC_seedCorEx.dscalar.nii")
```

```
## [1] TRUE
```